%% file: main.tex
\documentclass[conference,compsoc]{IEEEtran}
\IEEEoverridecommandlockouts

\ifCLASSOPTIONcompsoc
  \usepackage[nocompress]{cite}
\else
  \usepackage{cite}
\fi
\usepackage{graphicx}
\usepackage{amsmath}
\usepackage{algorithmic}
\usepackage{array}
\usepackage{colortbl}
\usepackage{amssymb}
\usepackage{pifont}
\usepackage{booktabs}
\usepackage{multirow}
\usepackage{tikz}
\usepackage{enumitem}
\usepackage{tcolorbox}
\usepackage{framed}

\ifCLASSOPTIONcompsoc
 \usepackage[caption=false,font=footnotesize,labelfont=sf,textfont=sf]{subfig}
\else
 \usepackage[caption=false,font=footnotesize]{subfig}
\fi
\usepackage{url}

\definecolor{formalshade}{rgb}{0.95,0.95,0.97}
\definecolor{darkblue}{rgb}{0.14,0.22,0.52}
\newenvironment{takeaway}{
\small

\MakeFramed{\advance\hsize-\width\FrameRestore}}
{\endMakeFramed}
\newcounter{takeaway}
\definecolor{darkorange}{rgb}{1.0, 0.549, 0.0}
\newenvironment{researchquestion}{
  \small

\MakeFramed{\advance\hsize-\width\FrameRestore}}
{\endMakeFramed}
\newcounter{researchquestion}
\definecolor{darkgreen}{rgb}{0.0, 0.549, 0.0}
\newenvironment{observation}{
  \small

\MakeFramed{\advance\hsize-\width\FrameRestore}}
{\endMakeFramed}
\newcounter{observation}
\newcommand{\ignore}[1]{}
\newcommand{\mysubref}[2]{\hyperref[#1]{\ref*{#1}(#2)}}

\newcommand{\whitecircle}{\tikz\draw[black] (0,0) circle (0.15cm);} 
\newcommand{\blackcircle}{\tikz\fill[black] (0,0) circle (0.15cm);} 

\newcommand{\revise}[1]{{\color{black}{#1}}}

\begin{document}

\title{SoK: Evaluating Jailbreak Guardrails for Large Language Models}

\author{
\IEEEauthorblockN{Xunguang Wang\IEEEauthorrefmark{1}, Zhenlan Ji\IEEEauthorrefmark{1}, Wenxuan Wang\IEEEauthorrefmark{2}, Zongjie Li\IEEEauthorrefmark{1}, Daoyuan Wu\IEEEauthorrefmark{3}, Shuai Wang\IEEEauthorrefmark{1}\IEEEauthorrefmark{4}\thanks{\IEEEauthorrefmark{4}Corresponding author.}}
\IEEEauthorblockA{\IEEEauthorrefmark{1}The Hong Kong University of Science and Technology, \IEEEauthorrefmark{2}Renmin University of China, \IEEEauthorrefmark{3}Lingnan University}
\IEEEauthorblockA{\texttt{\{xwanghm,zjiae,zligo,shuaiw\}@cse.ust.hk}, \texttt{wangwenxuan@ruc.edu.cn}, \texttt{daoyuanwu@ln.edu.hk}}
}

\maketitle

\begin{abstract}
  Large Language Models (LLMs) have achieved remarkable progress, but their deployment has exposed critical vulnerabilities, particularly to jailbreak attacks that circumvent safety alignments. Guardrails—external defense mechanisms that monitor and control LLM interactions—have emerged as a promising solution. However, the current landscape of LLM guardrails is fragmented, lacking a unified taxonomy and comprehensive evaluation framework. In this Systematization of Knowledge (SoK) paper, we present the first holistic analysis of jailbreak guardrails for LLMs. We propose a novel, multi-dimensional taxonomy that categorizes guardrails along six key dimensions, and introduce a Security-Efficiency-Utility evaluation framework to assess their practical effectiveness. Through extensive analysis and experiments, we identify the strengths and limitations of existing guardrail approaches, provide insights into optimizing their defense mechanisms, and explore their universality across attack types. Our work offers a structured foundation for future research and development, aiming to guide the principled advancement and deployment of robust LLM guardrails.
\end{abstract}

\IEEEpeerreviewmaketitle

\input{sec/introduction}
\input{sec/jailbreak}
\input{sec/criteria}
\input{sec/guardrail}
\input{sec/experiment}

\input{sec/discussion}

\section{Conclusion}
This SoK paper comprehensively addresses the fragmented landscape of LLM
jailbreak guardrails by introducing a novel multi-dimensional taxonomy and a SEU
measurement framework. Our findings highlight the strengths, limitations, and
interdependencies of existing defense mechanisms with a series of key insights.
This work forms a structured foundation to guide the principled advancement and
deployment of more robust LLM guardrails.

\section*{Acknowledgments}
We thank the reviewers and the shepherd for their
constructive comments. The HKUST authors are supported
in part by an RGC CRF grant under contract C6015-
23G and a research fund provided by HSBC.
Daoyuan Wu was partially supported by Lingnan Grant SUG-002/2526.

\bibliographystyle{IEEEtran}
\bibliography{llm, jailbreak, defense, guardrail, other}

\input{sec/appendix}
\input{sec/review}

\end{document}

%% file: sec/introduction.tex
\section{Introduction}
\label{sec:introduction}

Large Language Models (LLMs) have demonstrated remarkable capabilities across a wide range of applications, revolutionizing fields from natural language understanding to content generation~\cite{touvron2023llama2,Vicuna,Mistral7Bv02,llama3,Claude35sonnet,Qwen2.5}. However, their increasing sophistication and widespread adoption have also exposed significant vulnerabilities. A prominent concern is their susceptibility to \textit{jailbreak attacks}~\cite{GCG23,AutoDAN24}, where adversaries craft malicious inputs to bypass safety alignments and elicit harmful, biased, or unethical responses. The proliferation of such attacks underscores the urgent need for robust defense mechanisms. Among various defense strategies, \textit{guardrails}~\cite{NeMoGuardrails,LlamaGuard,SelfDefend} have emerged as a promising approach, aiming to monitor and control LLM interactions without altering the underlying model's weights or core functionalities.


Guardrail-based defenses offer a distinct advantage over other defense methods (e.g., tuning-based approaches~\cite{CAT24}) as they can effectively filter jailbreak attempts while preserving the integrity of the target LLM's original output capabilities. Despite their potential, the current landscape of LLM guardrails is characterized by \textit{siloed innovation}. Numerous research teams and organizations have proposed various guardrail solutions, often tailored to specific scenarios, attack vectors (e.g., focusing primarily on single-turn attacks), or proprietary systems. This ad-hoc development has resulted in a fragmented ecosystem of defense mechanisms, lacking a unified understanding or a systematic classification framework to position and compare these disparate efforts.

The absence of a systematic perspective contributes directly to a critical limitation in existing guardrails: a general lack of \textit{universality}. Many solutions are not readily adaptable across different LLMs, attack types, or deployment contexts. Furthermore, current evaluation practices for LLM guardrails often fall short of reflecting real-world operational constraints. Evaluations predominantly focus on raw defense efficacy against specific jailbreak benchmarks, frequently overlooking crucial factors such as \textit{computational cost} (e.g., inference latency, GPU resource consumption) and \textit{utility} (e.g., the rate of misclassifying benign prompts as malicious, thereby degrading user experience). This narrow evaluation scope hinders a comprehensive understanding of the practical trade-offs involved in deploying guardrails.

To address these critical gaps, this Systematization of Knowledge (SoK) paper provides the first comprehensive analysis and structuring of the rapidly evolving field of jailbreak guardrails for LLMs. We aim to consolidate the disparate research efforts, offering a clear and structured understanding of the current state-of-the-art. Our primary contributions are threefold: (1) we propose a novel, multi-dimensional taxonomy for classifying LLM guardrails, enabling a nuanced understanding of their design characteristics; (2) we introduce a holistic evaluation framework centered on the \textit{Security-Efficiency-Utility} trifecta, promoting more practical and comprehensive assessments; and (3) we conduct extensive analysis based on our framework, yielding valuable insights into the performance of existing guardrails and identifying promising avenues for future research.
Specifically, our contributions are as follows:
\begin{itemize}[leftmargin=*]
    \item \textbf{A Multi-Dimensional Guardrail Taxonomy:} We propose the first comprehensive taxonomy to categorize LLM guardrails along six critical dimensions:
    \begin{itemize}
        \item \textit{Intervention Stage}: Characterizing when the guardrail operates (Pre-processing, Intra-processing, or Post-processing of LLM interactions).
        \item \textit{Technical Paradigm}: Identifying the underlying mechanism (Rule-based, Model-based, or LLM-based).
        \item \textit{Safety Granularity}: Defining the scope of the guardrail detection (Token-level, Sequence-level, or Session-level).
        \item \textit{Reactivity}: Distinguishing between static (pre-defined) and dynamic (adaptive) defense strategies.
        \item \textit{Applicability}: Considering the guardrail's requirements regarding LLM access (White-box vs. Black-box).
        \item \textit{Interpretability}: Assessing the transparency of the guardrail's decision-making process. 
    \end{itemize}
    \item \textbf{A Security-Efficiency-Utility (SEU) Evaluation Framework:} We introduce a novel framework for evaluating guardrails that balances three crucial aspects:
    \begin{itemize}
        \item \textit{Security}: Measuring the defense performance against a diverse range of jailbreak attacks.
        \item \textit{Efficiency}: Quantifying the operational overhead, including inference delay and GPU memory consumption.
        \item \textit{Utility}: Assessing the impact on legitimate user interactions, primarily through the false positive rate on benign queries. 
    \end{itemize}
    \item \textbf{Experimental Findings and Optimization Insights:} We leverage our taxonomy and evaluation framework to analyze existing guardrails and explore future directions:
    \begin{itemize}
        \item We conduct a tri-objective (SEU) evaluation of mainstream guardrail methods to identify balanced solutions and those effective against diverse jailbreak categories.
        \item We investigate specific hypotheses, such as the efficacy of session-level guardrails against multi-turn attacks, the influence of intervention stage on latency, the impact of technical paradigms on resource consumption, and the relationship between safety granularity and utility.
        \item We explore the \textit{universality} of guardrails by assessing their performance against other attack modalities, such as prompt injection attacks.
    \end{itemize}
\end{itemize}

This SoK aims to provide researchers and practitioners with a clear roadmap for
understanding, developing, and deploying LLM jailbreak guardrails. By
systematizing existing knowledge and proposing a comprehensive evaluation
methodology, we hope to foster more principled advancements in this critical
area of LLM security. The code is available at \textcolor{blue}{\url{https://github.com/xunguangwang/SoK4JailbreakGuardrails}}.


%% file: sec/jailbreak.tex
\section{Jailbreak Attacks in LLMs}
\label{sec:jailbreak}

In this section, we first formally describe the jailbreak in LLMs and then introduce several typical jailbreak methods.

\noindent
\textbf{Jailbreak Formulation.}
The jailbreak phenomenon indicates that specific malicious instructions can bypass the safety mechanisms of LLMs, leading to the generation of harmful or unethical outputs. This is particularly concerning as it highlights the potential for adversaries to exploit vulnerabilities in LLMs to produce toxic or harmful content. The jailbreak process can be viewed as a two-step procedure: (1) crafting an adversarial prompt $P$ that elicits a harmful response from the LLM, and (2) evaluating the generated response $R$ against a predefined harmful objective $G$ using a classifier $\text{JUDGE}$. $\text{JUDGE}$ returns `True' if the generated response $R$ meets the harmful objective $G$, i.e., $\text{JUDGE}=\text{True}$, otherwise `False'.
Let $\mathcal{T}$ represent the LLM's vocabulary.
Formally, we can define the classifier $\text{JUDGE}: \mathcal{T}^\star \times \mathcal{T}^\star \to \{\text{True}, \text{False}\}$.
The adversary's goal is to maximize the probability of generating responses classified as satisfying the harmful objective $G$.
This can be expressed mathematically as:
\begin{equation}
    \sup_{P \in \mathcal{T}^\star} \Pr_{R \sim \text{LLM}(P)} [\text{JUDGE}(R, G) = \text{True}],
\end{equation}
where $\Pr$ denotes the probability, which accounts for the inherent stochasticity of the LLM's outputs when processing the input prompt $P$. The adversary iteratively refines prompts to identify those that maximize the likelihood of producing outputs deemed harmful by the classifier.

\noindent
\textbf{Existing Jailbreak Attacks.}
Jailbreak attacks can be broadly categorized into two types: single-turn and multi-turn jailbreaks. Single-turn jailbreaks involve crafting a single prompt to elicit harmful responses, while multi-turn jailbreaks exploit the interactive nature of LLMs by engaging in a dialogue with the model over multiple turns.
Due to the maturity of research on single-turn attacks and the relative scarcity of multi-turn attack studies, \revise{a further breakdown of the single-turn attack is necessary.}
\revise{Building on prior works~\cite{JBShield,SelfDefend} that categorize jailbreak attacks by their technical paradigms, we divide single-turn attacks into similar four types: manual, optimization-based, generation-based and implicit jailbreaks.}
\begin{itemize}[leftmargin=*]
    \item \textbf{Manual Jailbreaks.}~These attacks involve crafting prompts that exploit vulnerabilities in LLMs~\cite{Empirical23,Jailbroken23,ICD24,DAN24,MASTERKEY24}. Wei et al.~\cite{Jailbroken23} identified two key weaknesses—out-of-distribution inputs and conflicts between safety objectives and model capabilities—to inform prompt design. Deng et al.~\cite{MASTERKEY24} introduced AIM (Always Intelligent and Machiavellian), a proof-of-concept jailbreak prompt that served as a foundation for generating additional adversarial prompts. Shen et al.~\cite{DAN24} proposed JailbreakHub, a crowdsourcing framework for collecting diverse jailbreak prompts.
    \item \textbf{Optimization-based Jailbreaks.}~These methods iteratively refine adversarial prompts using techniques like gradient-based optimization or search strategies~\cite{GCG23, GCGPlus24, JSAA24, AutoDAN24, IGCG24}. GCG~\cite{GCG23} introduced a greedy coordinate gradient method to optimize adversarial suffixes, enabling transferable jailbreaks across models and prompts. Sitawarin et al.~\cite{GCGPlus24} extended this with GCG++, leveraging a proxy model to enhance optimization. Beyond gradient-based techniques, JSAA~\cite{JSAA24} employed random search for suffix optimization \revise{and BON~\cite{BON24} induces jailbreaks by repeatedly sampling augmentations of a harmful instruction until one succeeds}, while AutoDAN~\cite{AutoDAN24} used a hierarchical genetic algorithm to create human-readable jailbreak prompts. RLbreaker~\cite{RLbreaker24} utilized reinforcement learning to efficiently search for adversarial prompts, outperforming stochastic methods like JSAA and AutoDAN.
    \item \textbf{Generation-based Jailbreaks.}~These attacks use auxiliary LLMs to produce adversarial prompts~\cite{RedTeaming22, MASTERKEY24, PAIR23, TAP23, Advprompter24, AmpleGCG24}. PAIR~\cite{PAIR23} employs a feedback loop where the attacking LLM adjusts outputs based on the target LLM’s responses. Mehrotra et al.~\cite{TAP23} enhanced this approach using tree-of-thought reasoning~\cite{ToT}. LLM-Fuzzer~\cite{LLM-Fuzzer24} automates adversarial prompt generation by mutating human-written templates. Additionally, Advprompter~\cite{Advprompter24} trains a fine-tuned LLM to create both effective and human-readable adversarial prompts.
    \item \textbf{Implicit Jailbreaks.}~\revise{This class of attacks is derived from the Linguistics-based and Encoding-based jailbreak categories identified in JBShield~\cite{JBShield}. Given that both strategies conceal malicious intent within the query text to circumvent LLM safety mechanisms, they are collectively categorized as implicit jailbreaks
    .}
    For instance, Handa et al.~\cite{handa2024jailbreaking} demonstrated word substitution as a simple evasion method. DrAttack~\cite{DrAttack24} decomposes harmful prompts into smaller, less detectable sub-prompts. Puzzler~\cite{Puzzler24} embeds clues within queries to guide the LLM toward producing harmful outputs indirectly. Another approach involves translating harmful prompts into languages where LLM safety mechanisms are weaker~\cite{MultiJail23, yong2023low, DissectingMultilingual24, InvestigateMultilingual24, Jailbroken23, yuan2024cipherchat}. Deng et al.~\cite{MultiJail23} and Yong et al.~\cite{yong2023low} found that low-resource languages, such as Zulu, often exhibit less robust safety alignment. Obfuscation techniques~\cite{ComprehensiveJailbreak24}, including encoding or encrypting harmful prompts, further reduce LLM sensitivity to malicious inputs~\cite{Jailbroken23, yuan2024cipherchat}.
    \item \textbf{Multi-turn Jailbreaks.}~One multi-turn attack strategy is the fine-grained task decomposition, which decomposes the original malicious query into several less harmful sub-questions \cite{SimPO24, zhou2024speak,liu2024imposter}. While this decomposition strategy successfully circumvents current safety mechanisms, it may be easily mitigated by including these finer-grained harmful queries in safety training data. Alternatively, researchers propose to use human red teamers to expose vulnerabilities of LLMs against multi-turn attacks \cite{li2024llm}.
    Moreover, Yang et al.~\cite{yang2024chain} depend on the heuristics from \cite{PAIR23} and its seed examples to implement their attacks. Crescendo~\cite{Crescendo24} gradually steers benign initial queries towards more harmful topics. The implementation of Crescendo is based on the fixed and human-crafted seed instances, making it challenging to generate diverse and effective attacks. By contrast, ActorAttack~\cite{ActorAttack24} proposes to discover diverse attack clues inside the model’s prior knowledge. X-Teaming~\cite{X-Teaming25} achieves more effective and diverse multi-turn attacks by adaptive collaborative agents for planning, attack optimization, and verification.
\end{itemize}

\begin{figure}[t]
	\centering
	\includegraphics[width=0.95\columnwidth]{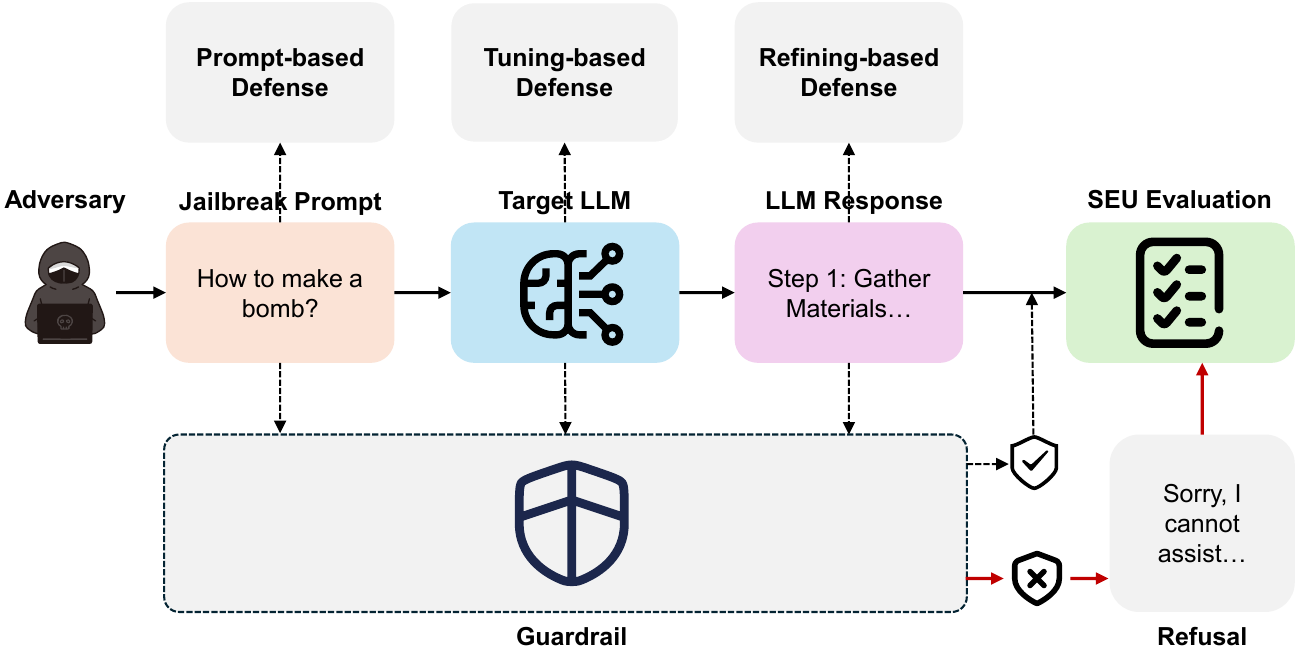}
	\caption{Illustration of a guardrail pipeline.}
	\label{fig:pipeline}
    \vspace{-2.0ex}
\end{figure}

%% file: sec/criteria.tex
\section{Definition, Taxonomy \& Evaluation}
\label{sec:criteria}

\revise{
\subsection{Jailbreak Defense}
\label{sec:jailbreak_defense}

Jailbreak defense comprises methods to protect LLMs from jailbreak attacks intended to elicit harmful or policy-violating content. We classify these defenses based on their point of application into four categories: prompt-based, tuning-based, refining-based defenses, and guardrails, which respectively focus on modifying prompts, optimizing model parameters, controlling model outputs, and external monitoring.

\textbf{Prompt-based defenses} alter the input prompt to enhance the LLM's adherence to safety guidelines~\cite{SelfReminder23,ICD24,RPO24,PAT24}. For instance, Self-Reminder~\cite{SelfReminder23} prepends ethical reminders, while ICD~\cite{ICD24} provides examples of harmful queries and safe responses.
%
\textbf{Tuning-based defenses} optimize the LLM's parameters to improve its inherent safety. This includes techniques like reinforcement learning from human feedback (RLHF)~\cite{InstructGPT22,ConstitutionalAI22}, machine unlearning to erase harmful knowledge~\cite{Eraser24}, model editing to patch vulnerabilities~\cite{LED24}, supervised fine-tuning on curated safety datasets~\cite{GoalPrioritization24, DeRTa25}, and adversarial training to bolster robustness against attacks~\cite{CAT24}.
%
\textbf{Refining-based defenses} steer the output generation process in real-time~\cite{RAIN23,SafeDecoding24,Aligner24}. For example, RAIN~\cite{RAIN23} uses token-level evaluation to guide rewinds for safer generation, whereas Aligner~\cite{Aligner24} employs a lightweight model to correct initial responses towards alignment.
%
\textbf{Guardrails} are external modules that monitor and control the interaction~\cite{NeMoGuardrails,LlamaGuard,WildGuard}. As illustrated in Figure~\ref{fig:pipeline}, they can operate on inputs (pre-processing), internal model states (intra-processing), or outputs (post-processing). Distinct from the other categories, guardrails do not modify the target LLM's inference; rather, they detect harmfulness and filter the interaction to prevent jailbreaks.

}

\subsection{Jailbreak Guardrail Definition}
\label{sec:jailbreak_guardrail_definition}

A ``Jailbreak Guardrail'' refers to a specialized security mechanism designed
for LLM systems, specifically to detect and prevent ``jailbreak''
attacks \cite{NeMoGuardrails,LlamaGuard,SelfDefend,WildGuard}. Such guardrails typically function as a
defensive layer, scrutinizing user inputs before they reach the LLM or vetting
the model's outputs before they are presented to the user. The primary objective
is to ensure that the LLM does not generate harmful, unethical, or
policy-violating content.

In the context of the jailbreak formulation introduced in Section~\ref{sec:jailbreak}, where an adversary crafts a prompt $P$ aiming to elicit a response $R = \text{LLM}(P)$ such that $\text{JUDGE}(R, G) = \text{True}$ (indicating a harmful outcome based on objective $G$), a jailbreak guardrail introduces an additional checkpoint. Let $\mathcal{G}_R$ denote the guardrail system, and $\text{Assess}(\mathcal{G}_R, X)$ be its assessment function, which returns $\textit{allow}$ if content $X$ (either $P$, $R$, or the internal feature $F$ of the LLM) is deemed permissible, and $\textit{block}$ otherwise. When $\textit{block}$ is executed, the final response $R$ is replaced by a safe response $R'$, such as ``\texttt{Sorry, I cannot assist with that}'', as shown in Figure~\ref{fig:pipeline}.

A jailbreak attack is considered successful in the presence of such a guardrail if, and only if, the protective mechanisms of both the target LLM (i.e., its inherent safety alignment) and the guardrail are circumvented. This means the guardrail must deem the interaction (either the input prompt or the generated output) as acceptable, while the target LLM still produces content classified as harmful.
More formally, if the guardrail inspects the input prompt $P$, a successful jailbreak occurs when:
\begin{equation}
\label{eq:jailbreak_guardrail_input}
\text{Assess}(\mathcal{G}_R, P) = \textit{allow} ~ \land ~ \text{JUDGE}(R, G) = \text{True}.
\end{equation}
Alternatively, if the guardrail inspects the model's internal features $F$ or output $R = \text{LLM}(P)$, a successful jailbreak is characterized respectively by:
\begin{equation}
\text{Assess}(\mathcal{G}_R, F) = \textit{allow} ~ \land ~ \text{JUDGE}(R, G) = \text{True},
\label{eq:jailbreak_guardrail_feature}
\end{equation}
\begin{equation}
\label{eq:jailbreak_guardrail_output}
\text{Assess}(\mathcal{G}_R, R) = \textit{allow} ~ \land ~ \text{JUDGE}(R, G) = \text{True}.
\end{equation}
This highlights that a successful adversary must not only craft a prompt that bypasses the LLM's internal safety measures but also deceive the guardrail into permitting the harmful interaction or content.

As jailbreak techniques become increasingly sophisticated and diverse (as noted in Section~\ref{sec:jailbreak}), these guardrails face mounting challenges. They must evolve beyond detecting overtly malicious requests to identify subtle and nuanced jailbreak patterns and adversarial manipulations. The continuous enhancement of jailbreak guardrails is therefore critical for improving the safety, security, and regulatory compliance of AI applications.


\subsection{Jailbreak Guardrail Taxonomy}
\label{sec:taxonomy}

This section categorizes existing guardrail approaches along several key
dimensions. Our taxonomy considers:
\begin{itemize}[leftmargin=*]
    \item \textbf{Intervention Stages}: This dimension delineates \textit{when} the guardrail operates within the LLM interaction pipeline—either at \textit{pre-processing} (before the input reaches the LLM), \textit{intra-processing} (during the LLM's inference), or \textit{post-processing} (after the LLM generates an output).
    \item \textbf{Technical Paradigms}: This refers to the underlying \textit{methodology} employed by the guardrail. Approaches are classified as \textit{rule-based} (relying on predefined rules or patterns), \textit{model-based} (using statistical models or classifiers), or \textit{LLM-based} (leveraging another LLM for analysis and decision-making).
    \item \textbf{Safety Granularity}: This specifies the \textit{level of detail} at which the safety analysis is performed. It can be \textit{token-level} (examining individual words or sub-word units), \textit{sequence-level} (evaluating entire prompts or responses), or \textit{session-level} (considering the context of the entire conversation history).
    \item \textbf{Reactiveness}: This dimension distinguishes how a guardrail responds to potentially harmful inputs. \textit{Static} defenses analyze inputs without modification, whereas \textit{dynamic} defenses actively alter inputs---for example, through mutation or perturbation---to neutralize adversarial properties while aiming to preserve overall semantic meaning.
    \item \textbf{Applicability}: This criterion assesses the guardrail's suitability for different LLM access models, with a particular emphasis on whether the mechanism can be effectively applied to \textit{black-box} LLMs (i.e., closed-source models or those accessed via remote APIs where internal states are not accessible).
    \item \textbf{Explainability}: This focuses on whether the guardrail method provides interpretable insights into its safety judgments or offers clear rationales for the decisions it makes.
\end{itemize}
This multi-faceted classification provides a comprehensive framework for understanding and navigating the landscape of LLM guardrails. We have comprehensively compiled existing works on jailbreak guardrails by this taxonomy, as summarized in Table~\ref{tab:summary}.

\subsection{Guardrail Evaluation Framework}
\label{sec:seu_framework}

To enable a comprehensive and practical assessment of LLM guardrails, we propose the \textit{Security-Efficiency-Utility} (SEU) Evaluation Framework. This framework is designed to capture the essential trade-offs involved in deploying guardrails in real-world LLM systems, moving beyond the narrow focus on raw defense efficacy. Below, we detail the three core dimensions of our framework and the specific metrics used for each.

\noindent
\textbf{Security: Defense Effectiveness.}
The primary objective of any guardrail is to enhance the security of LLM systems by mitigating jailbreak attacks. We evaluate defense effectiveness using two complementary metrics:
\begin{itemize}[leftmargin=*]
    \item \textbf{Attack Success Rate (ASR):} ASR measures the proportion of adversarial attempts that successfully bypass the guardrail and elicit harmful or unethical responses from the target LLM. Formally, it is defined as the percentage of attack queries for which the LLM system equipped with the guardrail fails to block or mitigate the attack. A lower ASR indicates a stronger defense.
    \item \textbf{Pass Guardrail Rate (PGR):} PGR measures the proportion of jailbreak attempts that successfully bypass the guardrail, indicating that the guardrail has classified the attempt as safe. For pre-processing and intra-processing guardrails, this refers to the proportion of malicious requests that the guardrail incorrectly identifies as benign. For post-processing guardrails, this refers to the proportion of instances where the guardrail fails to detect harmful content in the LLM's response to a jailbreak attempt. A lower PGR signifies a more effective guardrail in blocking attacks.
\end{itemize}

\noindent
\textbf{Efficiency: Computational Overhead.}
In practical deployments, the operational efficiency of guardrails is a critical consideration, as excessive overhead can degrade user experience and increase infrastructure costs. We assess efficiency along two axes:
\begin{itemize}[leftmargin=*]
    \item \textbf{Extra Delay:} This metric captures the additional response latency introduced by the guardrail. It is computed as the difference between the end-to-end response time of the guardrail + LLM system and that of the standalone target LLM. Formally,
    \begin{equation}
        \text{Extra Delay} = T_{\text{guardrail + LLM}} - T_{\text{LLM}},
    \end{equation}
    where $T_{\text{guardrail + LLM}}$ and $T_{\text{LLM}}$ denote the average response times with and without the guardrail, respectively.
    \item \textbf{GPU Memory Overhead:} This metric measures the increase in peak GPU memory consumption resulting from the integration of the guardrail. It is defined as the difference between the maximum GPU memory usage of the guardrail + LLM system and that of the target LLM alone:
    \begin{equation}
        \text{GPU Overhead} = M_{\text{guardrail + LLM}} - M_{\text{LLM}},
    \end{equation}
    where $M_{\text{guardrail + LLM}}$ and $M_{\text{LLM}}$ represent the peak GPU memory usage with and without the guardrail, respectively.
\end{itemize}

\noindent
\textbf{Utility: Impact on Benign Queries.}
A robust guardrail should not only block malicious inputs but also preserve the utility of the LLM for legitimate users. We quantify utility loss using the following metric:
\begin{itemize}[leftmargin=*]
    \item \textbf{False Positive Rate (FPR):} FPR measures the proportion of benign (non-malicious) queries that are incorrectly flagged or blocked by the guardrail. It is defined as the percentage of normal user queries that are misclassified as attacks. A lower FPR indicates better utility preservation, as the guardrail minimally disrupts legitimate interactions with the LLM.
\end{itemize}

\noindent
\textbf{Discussion.}
By jointly considering Security, Efficiency, and Utility, the SEU Evaluation Framework provides a holistic basis for comparing and optimizing LLM guardrails. This tri-objective perspective enables the identification of solutions that achieve a balanced trade-off, rather than excelling in only one dimension at the expense of others. In our experimental analysis (\S\ref{sec:experiment} \& \S\ref{sec:practical_implications}), we employ this framework to systematically evaluate mainstream guardrail methods, offering actionable insights for both researchers and practitioners.

\revise{
\noindent
\textbf{Alignment with Taxonomy Dimensions.}
The proposed taxonomy in \S\ref{sec:taxonomy} is intentionally structured to align with the empirical evaluation dimensions of the SEU framework, which form the core contribution of this work.
Particularly, we would analyze the first three taxonomy dimensions (Intervention Stage, Technical Paradigm, and Safety Granularity) in depth, because they have a direct, measurable impact on Security, Efficiency, and Utility, forming the foundation of our quantitative analysis.
In contrast, the latter three dimensions (i.e., Reactiveness, Applicability, and Explainability) are more qualitative in nature and represent critical future research directions for enhancing the adaptability, deployment flexibility, and transparency of guardrail systems. By focusing our empirical analysis on the dimensions most directly tied to operational performance, we provide a principled and actionable foundation for evaluating and optimizing jailbreak guardrails in practice.
}

%% file: sec/guardrail.tex
\section{Guardrail Analysis Based on Taxonomy}
\label{sec:guardrail}

\renewcommand{\arraystretch}{1.3} 
\setlength{\tabcolsep}{6pt} 

\begin{table*}[!ht]
    \centering
    \caption{Works on guardrails categorized by 6 dimensions. The black circle indicates the guardrail belongs to this dimension, and the white circle otherwise.
    }
    \vspace{-2.0ex}
    \setlength{\tabcolsep}{2pt}
    \resizebox{\textwidth}{!}{%
    \begin{tabular}{lc|ccc|ccc|ccc|cc|c|c}
    \toprule
    \multirow{2}{*}{\textbf{Paper}} & \multirow{2}{*}{\textbf{Venue}} & \multicolumn{3}{c|}{\textbf{Intervention Stages}} & \multicolumn{3}{c|}{\textbf{Technical Paradigms}} & \multicolumn{3}{c|}{\textbf{Safety Granularity}} & \multicolumn{2}{c|}{\textbf{Reactiveness}} & \multirow{2}{*}{\textbf{Applicability}} & \multirow{2}{*}{\textbf{Explainability}}  \\
    \cmidrule(lr){3-5} \cmidrule(lr){6-8} \cmidrule(lr){9-11} \cmidrule(lr){12-13}
     & & Pre-processing & Intra-processing & Post-processing & Rule & Model & LLM & Token & Sequence & Session & Static & Dynamic & & \\
    \midrule
    Perspective API \cite{PerspectiveAPI} & KDD'22 (2202.11176)          & \blackcircle & \whitecircle & \blackcircle & \whitecircle & \blackcircle & \whitecircle & \whitecircle & \blackcircle & \whitecircle & \blackcircle & \whitecircle & \blackcircle & \whitecircle \\
    OpenAI Moderation \cite{OpenAIModeration} & AAAI'23 (2208.03274)     & \blackcircle & \whitecircle & \blackcircle & \whitecircle & \blackcircle & \whitecircle & \whitecircle & \blackcircle & \whitecircle & \blackcircle & \whitecircle & \blackcircle & \whitecircle \\
    Self Defense~\cite{SelfDefense} & arXiv:2308.07308                   & \whitecircle & \whitecircle & \blackcircle & \whitecircle & \whitecircle & \blackcircle & \whitecircle & \blackcircle & \whitecircle & \blackcircle & \whitecircle & \blackcircle & \whitecircle \\
    Detecting Perplexity~\cite{DetectingPerplexity} & arXiv:2308.14132   & \blackcircle & \whitecircle & \whitecircle & \whitecircle & \blackcircle & \whitecircle & \whitecircle & \blackcircle & \whitecircle & \blackcircle & \whitecircle & \blackcircle & \blackcircle \\
    Perplexity Filter~\cite{PerplexityFilter} & arXiv:2309.00614         & \blackcircle & \whitecircle & \whitecircle & \whitecircle & \blackcircle & \whitecircle & \whitecircle & \blackcircle & \whitecircle & \blackcircle & \whitecircle & \blackcircle & \blackcircle \\
    erase-and-check~\cite{erase-and-check} & COLM'24 (2309.02705)        & \blackcircle & \whitecircle & \whitecircle & \whitecircle & \blackcircle & \blackcircle & \whitecircle & \blackcircle & \whitecircle & \whitecircle & \blackcircle & \blackcircle & \whitecircle \\
    SmoothLLM~\cite{SmoothLLM} & arXiv:2310.03684                        & \whitecircle & \whitecircle & \blackcircle & \blackcircle & \whitecircle & \whitecircle & \blackcircle & \whitecircle & \whitecircle & \whitecircle & \blackcircle & \blackcircle & \blackcircle \\
    NeMo Guardrails \cite{NeMoGuardrails} & EMNLP'23 (2310.10501)        & \blackcircle & \whitecircle & \blackcircle & \whitecircle & \whitecircle & \blackcircle & \whitecircle & \blackcircle & \whitecircle & \whitecircle & \blackcircle & \blackcircle & \whitecircle \\
    Llama Guard \cite{LlamaGuard} & arXiv:2312.06674                     & \blackcircle & \whitecircle & \blackcircle & \whitecircle & \whitecircle & \blackcircle & \whitecircle & \whitecircle & \blackcircle & \blackcircle & \whitecircle & \blackcircle & \blackcircle \\
    GradSafe~\cite{GradSafe} & ACL'24 (2402.13494)                       & \whitecircle & \blackcircle & \whitecircle & \whitecircle & \blackcircle & \whitecircle & \whitecircle & \blackcircle & \whitecircle & \blackcircle & \whitecircle & \whitecircle & \blackcircle \\
    SemanticSmooth \cite{SemanticSmooth} & arXiv:2402.16192              & \whitecircle & \whitecircle & \blackcircle & \whitecircle & \whitecircle & \blackcircle & \whitecircle & \blackcircle & \whitecircle & \whitecircle & \blackcircle & \blackcircle & \whitecircle \\
    LLMGuard~\cite{LLMGuard} & arXiv:2403.00826                          & \blackcircle & \whitecircle & \blackcircle & \blackcircle & \blackcircle & \whitecircle & \whitecircle & \blackcircle & \whitecircle & \blackcircle & \whitecircle & \blackcircle & \blackcircle \\
    Gradient Cuff~\cite{GradientCuff} & NeurIPS'24 (2403.00867)          & \whitecircle & \blackcircle & \whitecircle & \whitecircle & \blackcircle & \whitecircle & \whitecircle & \blackcircle & \whitecircle & \blackcircle & \whitecircle & \whitecircle & \blackcircle \\
    AutoDefense \cite{AutoDefense} & arXiv.2403.04783                    & \whitecircle & \whitecircle & \blackcircle & \whitecircle & \whitecircle & \blackcircle & \whitecircle & \blackcircle & \whitecircle & \blackcircle & \whitecircle & \blackcircle & \blackcircle \\
    RigorLLM~\cite{RigorLLM} & ICML'24 (2403.13031)                      & \blackcircle & \whitecircle & \whitecircle & \whitecircle & \blackcircle & \blackcircle & \whitecircle & \blackcircle & \whitecircle & \whitecircle & \blackcircle & \blackcircle & \blackcircle \\
    Aegis \cite{Aegis} & arXiv:2404.05993                                & \blackcircle & \whitecircle & \blackcircle & \whitecircle & \whitecircle & \blackcircle & \whitecircle & \whitecircle & \blackcircle & \blackcircle & \whitecircle & \blackcircle & \blackcircle \\
    LLMGuardrail~\cite{LLMGuardrail} & CCS'24 (2405.04160)               & \whitecircle & \blackcircle & \whitecircle & \whitecircle & \blackcircle & \whitecircle & \whitecircle & \blackcircle & \whitecircle & \blackcircle & \whitecircle & \whitecircle & \blackcircle \\
    RSAA~\cite{RSAA} & CAMLIS'24 (2406.03230)                            & \whitecircle & \blackcircle & \whitecircle & \whitecircle & \blackcircle & \whitecircle & \whitecircle & \blackcircle & \whitecircle & \blackcircle & \whitecircle & \whitecircle & \whitecircle \\
    Circuit Breaking~\cite{CircuitBreaking} & NeurIPS'24 (2406.04313)    & \whitecircle & \blackcircle & \whitecircle & \whitecircle & \blackcircle & \whitecircle & \whitecircle & \blackcircle & \whitecircle & \blackcircle & \whitecircle & \whitecircle & \whitecircle \\
    SelfDefend~\cite{SelfDefend} & USENIX Security'25 (2406.05498)       & \blackcircle & \whitecircle & \whitecircle & \whitecircle & \whitecircle & \blackcircle & \blackcircle & \blackcircle & \whitecircle & \blackcircle & \whitecircle & \blackcircle & \blackcircle \\
    GuardAgent~\cite{GuardAgent} & ICML'25 (2406.09187)       & \blackcircle & \whitecircle & \whitecircle & \whitecircle & \whitecircle & \blackcircle & \whitecircle & \blackcircle & \whitecircle & \blackcircle & \whitecircle & \blackcircle & \blackcircle \\
    WildGuard \cite{WildGuard} & NeurIPS'24 (2406.18495)                 & \blackcircle & \whitecircle & \blackcircle & \whitecircle & \whitecircle & \blackcircle & \whitecircle & \whitecircle & \blackcircle & \blackcircle & \whitecircle & \blackcircle & \whitecircle \\
    $R^{2}$-Guard \cite{R2Guard} & ICLR'25 (2407.05557)                  & \blackcircle & \whitecircle & \blackcircle & \whitecircle & \blackcircle & \whitecircle & \whitecircle & \blackcircle & \whitecircle & \blackcircle & \whitecircle & \blackcircle & \blackcircle \\
    Prompt Guard~\cite{PromptGuard,GAP} & Hugging Face (22 July 2024)    & \blackcircle & \whitecircle & \whitecircle & \whitecircle & \blackcircle & \whitecircle & \whitecircle & \whitecircle & \blackcircle & \blackcircle & \whitecircle & \blackcircle & \whitecircle \\
    PrimeGuard~\cite{PrimeGuard} & arXiv:2407.16318                      & \blackcircle & \whitecircle & \whitecircle & \whitecircle & \whitecircle & \blackcircle & \whitecircle & \blackcircle & \whitecircle & \blackcircle & \whitecircle & \blackcircle & \blackcircle \\
    ShieldGemma \cite{ShieldGemma} & arXiv:2407.21772                    & \blackcircle & \whitecircle & \blackcircle & \whitecircle & \whitecircle & \blackcircle & \whitecircle & \whitecircle & \blackcircle & \blackcircle & \whitecircle & \blackcircle & \blackcircle \\
    Adaptive Guardrail \cite{AdaptiveGuardrail} & arXiv:2408.08959       & \blackcircle & \whitecircle & \whitecircle & \whitecircle & \blackcircle & \whitecircle & \whitecircle & \whitecircle & \blackcircle & \blackcircle & \whitecircle & \blackcircle & \whitecircle \\
    EEG-Defender~\cite{EEG-Defender} & ICONIP'25 (2408.11308)        & \whitecircle & \blackcircle & \whitecircle & \whitecircle & \blackcircle & \whitecircle & \whitecircle & \blackcircle & \whitecircle & \blackcircle & \whitecircle & \whitecircle & \blackcircle \\
    HSF~\cite{HSF} & WWW'25 (2409.03788)                      & \whitecircle & \blackcircle & \whitecircle & \whitecircle & \blackcircle & \whitecircle & \whitecircle & \blackcircle & \whitecircle & \blackcircle & \whitecircle & \whitecircle & \whitecircle \\
    MoJE~\cite{MoJE} & AIES'24 (2409.17699)                              & \blackcircle & \whitecircle & \whitecircle & \whitecircle & \blackcircle & \whitecircle & \whitecircle & \blackcircle & \whitecircle & \blackcircle & \whitecircle & \blackcircle & \whitecircle \\
    Rapid Response~\cite{RapidResponse} & arXiv:2411.07494               & \blackcircle & \whitecircle & \whitecircle & \blackcircle & \blackcircle & \blackcircle & \whitecircle & \blackcircle & \whitecircle & \whitecircle & \blackcircle & \blackcircle & \whitecircle \\
    Pretrained Embeddings~\cite{PretrainedEmbeddings} & arXiv:2412.01547 & \blackcircle & \whitecircle & \whitecircle & \whitecircle & \blackcircle & \whitecircle & \whitecircle & \blackcircle & \whitecircle & \blackcircle & \whitecircle & \blackcircle & \whitecircle \\
    Token Highlighter~\cite{TokenHighlighter} & AAAI'25 (2412.18171)     & \whitecircle & \blackcircle & \whitecircle & \whitecircle & \blackcircle & \whitecircle & \blackcircle & \whitecircle & \whitecircle & \whitecircle & \blackcircle & \whitecircle & \whitecircle \\
    Aegis2.0 \cite{Aegis2.0} & arXiv:2501.09004                          & \blackcircle & \whitecircle & \blackcircle & \whitecircle & \whitecircle & \blackcircle & \whitecircle & \whitecircle & \blackcircle & \blackcircle & \whitecircle & \blackcircle & \blackcircle \\
    COT Fine-Tuning~\cite{COTFine-Tuning} & arXiv:2501.13080             & \blackcircle & \whitecircle & \whitecircle & \whitecircle & \whitecircle & \blackcircle & \whitecircle & \blackcircle & \whitecircle & \blackcircle & \whitecircle & \blackcircle & \blackcircle \\
    GuardReasoner \cite{GuardReasoner} & arXiv:2501.18492                & \blackcircle & \whitecircle & \blackcircle & \whitecircle & \whitecircle & \blackcircle & \whitecircle & \whitecircle & \blackcircle & \blackcircle & \whitecircle & \blackcircle & \blackcircle \\
    Constitutional Classifiers~\cite{ConstiCls} & arXiv:2501.18837       & \blackcircle & \whitecircle & \blackcircle & \whitecircle & \whitecircle & \blackcircle & \whitecircle & \blackcircle & \whitecircle & \blackcircle & \whitecircle & \blackcircle & \whitecircle \\
    JBShield~\cite{JBShield} & USENIX Security'25 (2502.07557)           & \whitecircle & \blackcircle & \whitecircle & \whitecircle & \blackcircle & \whitecircle & \whitecircle & \blackcircle & \whitecircle & \blackcircle & \whitecircle & \whitecircle & \blackcircle \\
    EDDF~\cite{EDDF} & ACL Fingdings'25 (2502.19041)       & \blackcircle & \whitecircle & \whitecircle & \whitecircle & \blackcircle & \whitecircle & \whitecircle & \blackcircle & \whitecircle & \blackcircle & \whitecircle & \blackcircle & \blackcircle \\
    CURVALID~\cite{CURVALID} & arXiv:2503.03502                          & \blackcircle & \whitecircle & \whitecircle & \whitecircle & \blackcircle & \whitecircle & \blackcircle & \blackcircle & \whitecircle & \blackcircle & \whitecircle & \blackcircle & \blackcircle \\
    MirrorShield~\cite{MirrorShield} & arXiv:2503.12931                    & \whitecircle & \blackcircle & \whitecircle & \whitecircle & \blackcircle & \whitecircle & \whitecircle & \blackcircle & \whitecircle & \blackcircle & \whitecircle & \blackcircle & \blackcircle \\
    JailGuard~\cite{JailGuard} & TOSEM'25 (19 March 2025)                & \whitecircle & \whitecircle & \blackcircle & \whitecircle & \blackcircle & \whitecircle & \whitecircle & \blackcircle & \whitecircle & \whitecircle & \blackcircle & \blackcircle & \blackcircle \\
    X-Guard~\cite{X-Guard} & arXiv:2504.08848                            & \blackcircle & \whitecircle & \blackcircle & \whitecircle & \whitecircle & \blackcircle & \whitecircle & \blackcircle & \whitecircle & \whitecircle & \blackcircle & \blackcircle & \blackcircle \\
    Continuous Detector~\cite{JailbreaksOverTime} & arXiv:2504.19440                                  & \blackcircle & \whitecircle & \blackcircle & \whitecircle & \whitecircle & \blackcircle & \whitecircle & \whitecircle & \blackcircle & \blackcircle & \whitecircle & \blackcircle & \whitecircle \\
    Active Monitoring~\cite{JailbreaksOverTime} & arXiv:2504.19440                                  & \blackcircle & \whitecircle & \blackcircle & \whitecircle & \whitecircle & \blackcircle & \whitecircle & \whitecircle & \blackcircle & \blackcircle & \whitecircle & \blackcircle & \blackcircle \\
    \bottomrule
    \end{tabular}
    }
\label{tab:summary}
\vspace{-3.0ex}
\end{table*}

\subsection{Intervention Stages}
Guardrail mechanisms can be deployed at different stages of the LLM interaction pipeline, including pre-processing, intra-processing, and post-processing. Each stage serves a distinct purpose in identifying and mitigating jailbreak attempts:

\noindent
\textbf{Pre-processing Guardrails.}
These mechanisms operate on user inputs before they reach the target LLM, functioning as the first line of defense against jailbreak attempts. Pre-processing guardrails typically employ detection algorithms to identify potentially harmful prompts and then block them entirely.
These guards are particularly valuable for their ability to prevent harmful prompts from ever reaching the model, thus conserving computational resources and reducing potential risks.

Early methods, such as Detecting Perplexity \cite{DetectingPerplexity} and Perplexity Filter \cite{PerplexityFilter}, compute the perplexity of input prompts to detect potential adversarial inputs. However, this approach is limited to GCG \cite{GCG23,GCGPlus24,AmpleGCG24} attacks with unreadable adversarial suffixes amplifying the perplexity.

A more direct approach is to identify the semantic harmfulness of input sequences. Some methods focus on directly identifying toxic phrases or excerpts within the input text \cite{SelfDefend,WildGuard,RapidResponse}, while others assess the overall semantic harmfulness of the entire input \cite{PerspectiveAPI,OpenAIModeration,erase-and-check,LlamaGuard,RigorLLM,Aegis,WildGuard,R2Guard,PromptGuard,PrimeGuard,ShieldGemma,AdaptiveGuardrail,MoJE,PretrainedEmbeddings,Aegis2.0,JailbreaksOverTime}. For instance, PromptGuard \cite{PromptGuard} and OpenAI Moderation \cite{OpenAIModeration} fine-tune pre-trained classifiers to assess the safety of input prompts. However, pre-processing guardrails may struggle with novel jailbreak techniques that do not exhibit clear patterns, e.g., implicit attack DrAttack~\cite{DrAttack24} conceals malicious content within benign-looking prompts.

A more fundamental approach is to analyze the true intent of the query to filter out jailbreak requests, based on the premise that jailbreak attempts always involve malicious output targets. Leveraging the powerful language understanding capabilities of LLMs, we can directly utilize LLMs to identify the real intentions of requests to determine whether they are jailbreak attempts~\cite{SelfDefend,GuardAgent,EDDF,JailbreaksOverTime}. For example, SelfDefend \cite{SelfDefend} with the intent prompt first summarizes the input intention and then assesses whether it constitutes a jailbreak request. Recently, some studies have employed LLM reasoning capabilities to analyze input intent \cite{COTFine-Tuning,GuardReasoner,X-Guard} before the safety judgments. For instance, X-Guard \cite{GuardReasoner} employs deep thinking to evaluate potential harms.

\begin{takeaway}
\addtocounter{takeaway}{1}
\noindent\textbf{Summary \thetakeaway: }
\textit{Pre-processing guardrails are the first line of defense against jailbreak attempts, operating on user inputs before they reach the target LLM. They have evolved from simple perplexity detection to semantic harmfulness identification and, most recently, to LLM-based reasoning for analyzing input intent. This evolution is driven by the need to address increasingly sophisticated and covert attack methods.}
\end{takeaway}

\noindent
\textbf{Intra-processing Guardrails.}
These guardrails operate during the LLM's inference process, analyzing internal model features or gradients to detect potential jailbreak attempts. Unlike pre-processing methods, intra-processing guardrails can observe how the model processes inputs internally, providing deeper insights into potential vulnerabilities.

On one hand, intra-processing guardrails rely on gradient information to
identify potential jailbreak attempts. These methods analyze the gradients of
the model's inputs or parameters during inference to identify unusual patterns
or anomalies that may indicate adversarial inputs. For example,
GradSafe~\cite{GradSafe} computes the similarity between the input's gradient
w.r.t. the safety-critical parameters and the unsafe reference gradients.
Gradient Cuff~\cite{GradientCuff} compares the gradient norm of refusal loss
w.r.t. the query prompt with a threshold. Token
Highlighter~\cite{TokenHighlighter} uses the gradient norm of the affirmation
loss for each token in the user query to locate the jailbreak-critical tokens.

On the other hand, intra-processing guardrails can analyze the model's internal
states for jailbreak detection
\cite{LLMGuardrail,RSAA,CircuitBreaking,EEG-Defender,HSF,JBShield}. These
methods leverage the model's hidden states or other internal representations, to
identify patterns indicative of jailbreak attempts. For example, Circuit
Breaking \cite{CircuitBreaking} interrupts the LLM to output harmful content
when harmful states are detected. JBShield~\cite{JBShield} analyzes the
differences of the LLM's internal states between the jailbreak prompts and the
benign queries. These approaches can provide more nuanced insights into the
model's behavior and vulnerabilities, enabling more effective detection of
sophisticated jailbreak techniques.
However, these approaches typically require white-box access to the target model, which limits their applicability to open-source LLMs or scenarios where model internals are accessible.

\begin{takeaway}
\addtocounter{takeaway}{1}
\noindent\textbf{Summary \thetakeaway: }
\textit{Intra-processing guardrails operate during the LLM's inference process, analyzing internal model features or gradients to detect potential jailbreak attempts. They provide deeper insights into vulnerabilities but require white-box access to the model, limiting their applicability.}
\end{takeaway}

\noindent
\textbf{Post-processing Guardrails.}
These mechanisms evaluate the LLM's generated outputs to identify and filter harmful content. As jailbreak attacks inherently aim to produce harmful outputs, post-processing guardrails serve as a crucial last line of defense. This ensures that even if malicious prompts circumvent earlier detection stages, their resultant outputs can still be intercepted.

Given that post-processing guardrails scrutinize the LLM's generated outputs, a primary strategy involves the direct detection of harmfulness within these responses. The most elementary of these methods employ keyword-based detectors to assess the safety of the LLM's outputs, primarily focusing on determining its jailbroken state \cite{SmoothLLM,LLMGuard}.
Building upon this foundational technique, a more sophisticated approach involves training dedicated classifiers to distinguish between harmful and harmless responses \cite{PerspectiveAPI,OpenAIModeration,LLMGuard,R2Guard}. For instance, initiatives like the Perspective API~\cite{PerspectiveAPI} and OpenAI Moderation~\cite{OpenAIModeration} have developed transformer-based classifiers engineered to predict the probability of harmful content appearing in an LLM's response. Similarly, $R^{2}$-Guard~\cite{R2Guard} embeds safety knowledge into probabilistic graphical models, enabling the computation of unsafe probabilities for any given LLM outputs.
Elevating this classification paradigm further, albeit with increased computational demands, some techniques leverage the reasoning capabilities of other LLMs to assess response safety \cite{SelfDefense,NeMoGuardrails,LlamaGuard,SemanticSmooth,AutoDefense,Aegis,WildGuard,ShieldGemma,Aegis2.0,GuardReasoner,ConstiCls,X-Guard,JailbreaksOverTime}. Self Defense~\cite{SelfDefense}, for example, filters harmful content by querying an LLM about the harmfulness of the initial response. Llama Guard~\cite{LlamaGuard} takes a more contextual approach by considering the input prompt in conjunction with the output to determine the risk category of the response. Progressing towards even more thorough analysis, GuardReasoner~\cite{GuardReasoner} and X-Guard~\cite{X-Guard} employ chain-of-thought reasoning before rendering a safety judgment on the LLM's output.

Beyond directly assessing the semantic harmfulness of the response, an alternative category of methods ingeniously leverages the discrepancies in outputs that arise from disrupting the adversarial characteristics of the initial jailbreak prompt \cite{SmoothLLM,SemanticSmooth,MirrorShield,JailGuard}. SmoothLLM~\cite{SmoothLLM}, for instance, operates by randomly perturbing or permuting multiple copies of a given input prompt to generate a set of responses; the safety of the original request is then determined by a voting mechanism based on these perturbed responses. Advancing this concept, SemanticSmooth~\cite{SemanticSmooth} and JailGuard~\cite{JailGuard} implement more complex mutations than the simple character-level alterations used by SmoothLLM, such as paraphrasing the prompt or translating it into other languages. In a similar vein, MirrorShield~\cite{MirrorShield} generates \revise{``mirror"} prompts that aim to preserve the syntactic structure of the input while ensuring its semantic safety.
These mutation-based methods capitalize on the inherent properties of jailbreak prompts to identify adversarial inputs, rendering them particularly effective against sophisticated attacks. Nevertheless, they may incur additional computational overhead due to the requirement for numerous response evaluations or intricate input transformations.

Although detecting the output of LLMs may appear more straightforward than deciphering ambiguous prompts and internal features, later methodologies will integrate the input prompts to thoroughly evaluate the safety of the query. However, post-processing safeguards may incur more latency than other paradigms due to the requirement of awaiting the LLM's response. Furthermore, mutation-based techniques that mandate multiple response evaluations are suspected to exacerbate this latency.

\begin{takeaway}
    \addtocounter{takeaway}{1}
    \noindent\textbf{Summary \thetakeaway: }
    \textit{Post-processing guardrails, by operating on the LLM's generated outputs to identify and filter harmful content, act as an essential safeguard. They intercept potentially harmful outputs that bypass earlier detection stages. However, over-reliance on the nature of responses may cause noticeable delay, particularly when employing mutation-based techniques that need multiple evaluations.}
\end{takeaway}

Drawing upon a classification by intervention stages, our analysis reveals a
critical gap in the current literature, which motivates the RQ below. As no
prior work has systematically investigated this specific dimension, we undertake
a thorough examination in this paper.

\begin{researchquestion}
    \addtocounter{researchquestion}{1}
    \noindent\textbf{RQ \theresearchquestion: } \textit{Pre-processing
    guardrails can reject harmful inputs before they reach the target LLM,
    intra-processing mechanisms operate concurrently during LLM inference, and
    post-processing techniques must await LLM's outputs. This raises a pertinent
    question: To what extent does the specific intervention stage of a
    guardrail, be it pre-processing, intra-processing, or post-processing,
    influence overall response latency?}
    \vspace{-1.0ex}
\end{researchquestion}

\subsection{Technical Paradigms}
Guardrail mechanisms employ diverse technical approaches to detect and mitigate jailbreak attempts, including rule-based, model-based, and LLM-based approaches.

\noindent
\textbf{Rule-based Guardrails.}
These guardrails operate by employing predefined rules, patterns, or heuristics to detect potentially harmful inputs or outputs of LLMs. A typical rule-based approach includes utilizing keywords or regular expressions to identify specific patterns tied to harmful content. For instance, SmoothLLM, as referenced in \cite{SmoothLLM}, leverages keyword-based detectors to assess the safety of the LLM’s outputs, primarily to determine its jailbroken state. Similarly, the PII Detector mentioned in \cite{LLMGuard} uses regular expressions to identify personally identifiable information, such as phone numbers and emails. This method mirrors the approach taken by the baseline Regex in \cite{RapidResponse}, which also utilizes regular expressions to mitigate jailbreak attacks.

Transitioning from specific examples to an evaluation of their effectiveness, it is evident that while these methods benefit from straightforward and transparent pattern matching—attributes that contribute to their computational efficiency and interpretability—their reliance on predefined patterns can be a significant drawback. Specifically, these rule-based systems may falter when encountering novel jailbreak techniques that deviate from known patterns which inherently limit their capability to combat more sophisticated attacks.

\begin{takeaway}
\addtocounter{takeaway}{1}
\noindent\textbf{Summary \thetakeaway: }
\textit{Rule-based guardrails, while beneficial for their computational efficiency and ease of interpretation, face challenges when dealing with innovative jailbreak techniques that do not match existing predefined patterns.}
\vspace{-1.0ex}
\end{takeaway}

\noindent
\textbf{Model-based Guardrails.}
These guardrails adopt classifiers or statistical characteristics to distinguish between benign and harmful queries. Model-based approaches can capture more complex patterns than rule-based methods, enabling them to generalize better to novel jailbreak attempts.
Learning a text-based classifier is a common approach for jailbreak detection.
On one hand, we can use traditional machine learning models as the classifiers \cite{RigorLLM,RSAA,R2Guard,HSF,MoJE,RapidResponse,PretrainedEmbeddings}. For instance, K-Nearest Neighbors (KNN) in RigorLLM~\cite{RigorLLM}, LightGBM in RSAA~\cite{RSAA} and Random Forest in PretrainedEmbeddings~\cite{PretrainedEmbeddings}.
On the other hand, neural networks are also widely applied for the safety classification \cite{PerspectiveAPI,OpenAIModeration,erase-and-check,LLMGuard,LLMGuardrail,PromptGuard,HSF,RapidResponse,CURVALID}. For example, HSF~\cite{HSF} and CURVALID~\cite{CURVALID} use a simple Multilayer Perceptron (MLP) as the classifier. PromptGuard~\cite{PromptGuard} and erase-and-check~\cite{erase-and-check} fine-tune the pre-trained model (i.e, mDeBERTa and DistilBERT, respectively) to distinguish the safe and unsafe inputs.
Besides, other methods used statistical characteristics to design their own algorithms on safety distinguish \cite{GradSafe,GradientCuff,AdaptiveGuardrail,EEG-Defender,TokenHighlighter,JBShield,EDDF,MirrorShield,JailGuard}. Detecting Perplexity~\cite{DetectingPerplexity} and Perplexity Filter~\cite{PerplexityFilter} classify the input as a jailbreak request if the perplexity is higher than a threshold. Gradient discrepancies between safe prompts and adversarial prompts are employed in GradSafe~\cite{GradSafe}, GradientCuff~\cite{GradientCuff} and TokenHighlighter~\cite{TokenHighlighter}. JailGuard~\cite{JailGuard} identifies the jailbroken state of responses by computing their KL-divergence. EEG-Defender~\cite{EEG-Defender}, JBShield~\cite{JBShield} and MirrorShield~\cite{MirrorShield} take the model's internal feature similarities between the input and the jailbreak prompt as judgment basis.

The essence of model-based guardrails is to find a classification standard in distinguishing the benign and harmful requests, whether to learn a classifier or design a statistical algorithm. Compared with rule-based methods, model-based approaches can capture more complex patterns and generalize better to novel jailbreak attempts. They can also adapt to evolving threats by retraining or fine-tuning the classifiers. However, these methods typically require substantial training data and computational resources, especially when using deep learning models.

\begin{takeaway}
\addtocounter{takeaway}{1}
\noindent\textbf{Summary \thetakeaway: }
\textit{Model-based guardrails, by adopting classifiers or statistical characteristics to distinguish between benign and harmful queries, can capture more complex patterns than rule-based methods, enabling them to generalize better to novel jailbreak attempts. However, these methods typically require substantial training data and computational resources, especially when using deep learning models.}
\end{takeaway}

\begin{observation}
    \addtocounter{observation}{1}
    \noindent\textbf{Observation: }
    \textit{Intra-processing guardrails are basically model-based guardrails, which use LLM's internal features to detect potential jailbreak attempts. This is because model-based methods analyze the features and build classifiers instead of using simple character matching or a more complex LLM.}
\end{observation}

\noindent
\textbf{LLM-based Guardrails.}
LLM-based guardrails represent a sophisticated approach to security, harnessing the inherent inferring capabilities of LLMs themselves to identify and counteract jailbreak attempts. Within this paradigm, research has progressed through distinct phases, each characterized by evolving methodologies.

Initially, methods tended to focus on directly determining the harmfulness of a request or providing a summary analysis after the judgment. For example, Self Defense~\cite{SelfDefense} directly employs the target LLM to assess the safety of its own generated responses and subsequently furnish an explanation for its findings. In a similar vein, Llama Guard~\cite{LlamaGuard} operates by first identifying an unsafe text and then assigning it to a harmfulness category. Complementing these approaches, WildGuard~\cite{WildGuard} offers a multi-faceted assessment, simultaneously reporting the harmfulness status of the input prompt, the generated response, and whether the response was ultimately refused.

More recently, however, there has been a discernible shift towards methodologies that conduct a more detailed, upfront analysis before arriving at a safety judgment. Illustrating this trend, SelfDefend~\cite{SelfDefend} first summarizes the underlying intention of the input and then assesses whether this intention constitutes a jailbreak request. Building upon this principle of preliminary in-depth analysis, both GuardReasoner~\cite{GuardReasoner} and X-Guard~\cite{X-Guard} employ chain-of-thought reasoning. This allows them to meticulously trace and analyze potential harms associated with a query, culminating in a final safety judgment.

Undeniably, the strength of these LLM-driven techniques lies in the excellent language understanding intrinsic to the models themselves. As a result, these approaches demonstrate considerable efficacy in detecting a diverse range of jailbreak attempts and notably improve the explainability of the safety judgments they provide.
Nevertheless, a crucial trade-off exists. While effective and explainable, these advanced guardrails may introduce substantially more computational overhead when compared to rule-based and model-based techniques.

\begin{takeaway}
    \addtocounter{takeaway}{1}
    \noindent\textbf{Summary \thetakeaway: }
    \textit{Employing the reasoning capabilities of LLMs, LLM-based guardrails not only detect and mitigate jailbreak attempts effectively but also improve the explainability of safety judgments. Nonetheless, they introduce significantly greater computational overhead than traditional rule-based and model-based methods.}
\end{takeaway}

We now present one RQ that focuses on the cost of LLM-based guardrails, which is
a crucial aspect of their practical deployment. This RQ is particularly relevant
given the increasing complexity and resource demands of LLM-based approaches,
especially in environments with limited computational resources.

\begin{researchquestion}
    \addtocounter{researchquestion}{1}
    \noindent\textbf{RQ \theresearchquestion: } \textit{Given that rule-based,
    model-based, and LLM-based guardrails inherently possess different levels of
    computational complexity and resource requirements, a significant practical
    question emerges: To what extent does the choice of technical paradigm
    directly influence the GPU memory footprint of LLM guardrail mechanisms
    during their operational deployment?}
\end{researchquestion}

\subsection{Safety Granularity}
Guardrail mechanisms can operate at 3 different levels of detection granularity: token-level for individual words or tokens, sequence-level for an entire prompt or response, and session-level for entire conversation sessions.

\noindent
\textbf{Token-level Guardrails.}
These guardrails analyze individual tokens or small token groups to identify potentially harmful elements within inputs or outputs. Token-level approaches can pinpoint specific problematic components within a text, enabling more precise interventions. For instance, Token Highlighter~\cite{TokenHighlighter} identifies specific tokens that contribute to harmful outputs. These fine-grained approaches enable targeted interventions but may miss harmful content that emerges from the broader context rather than specific tokens.

\noindent
\textbf{Sequence-level Guardrails.}
These guardrails evaluate entire prompts or responses as cohesive units, considering the overall semantic meaning rather than individual components. Sequence-level approaches can capture harmful content that emerges from the interaction between different parts of a text. For example, Llama Guard~\cite{LlamaGuard} and ShieldGemma~\cite{ShieldGemma} assess the holistic safety of the prompt sequence, while Constitutional Classifiers~\cite{ConstiCls} evaluate outputs against predefined safety principles. These approaches can better capture contextual harms but may provide less granular insights into specific problematic elements.

\noindent
\textbf{Session-level Guardrails.}
These guardrails monitor entire conversation sessions, tracking the evolution of dialogue across multiple turns to identify potential jailbreak attempts that unfold gradually. Session-level approaches can detect sophisticated multi-turn attacks that might appear benign when individual messages are analyzed in isolation. For instance, Adaptive Guardrail~\cite{AdaptiveGuardrail} maintains conversation state to identify harmful patterns across turns. These comprehensive approaches are particularly valuable against advanced jailbreak techniques that exploit the sequential nature of conversations but typically require more complex implementation and greater computational resources.
We now present two RQs that explore the impact of safety granularity on
the effectiveness and utility of guardrail mechanisms.

\begin{researchquestion}
    \addtocounter{researchquestion}{1}
    \noindent\textbf{RQ \theresearchquestion: }
    \textit{To what extent are current session-level guardrails truly effective in defending against sophisticated multi-turn jailbreak attacks?}
\end{researchquestion}

\begin{researchquestion}
    \addtocounter{researchquestion}{1}
    \noindent\textbf{RQ \theresearchquestion: }
    \textit{How does the choice of safety granularity (i.e., token, sequence, or session-level) impact the utility of LLMs when implementing guardrail mechanisms?}
\end{researchquestion}

\revise{
\subsection{Reactiveness}
\textbf{Static Guardrails} operate by analyzing inputs or outputs of LLMs without modifying them. For example, OpenAI Moderation~\cite{OpenAIModeration} uses a fine-tuned model to classify input prompts into safety categories without altering the original query. 
In contrast, \textbf{Dynamic Guardrails} actively modify inputs or outputs to neutralize adversarial properties while preserving semantic meaning, which is beneficial for better identification of jailbreak attacks.
The modification on the input prompt or the output response can be divided into character-level, token-level, word-level, and sentence-level changes.

\textit{Character-level} changes involve randomly introducing and altering characters within the text input, such as typos, inserting new characters or character swaps, to disrupt adversarial prompts without changing the overall meaning~\cite{SmoothLLM,SemanticSmooth}. A prominent example is SmoothLLM~\cite{SmoothLLM}, which perturbs multiple copies of the input prompt through character-level changes and uses a voting mechanism to determine safety.
\textit{Token-level} changes involve substituting, inserting, or deleting specific tokens to neutralize adversarial prompts. For instance, erase-and-check~\cite{erase-and-check} removes potentially harmful tokens from the input prompt to construct multiple sanitized versions for safety voting like SmoothLLM.
\textit{Word-level} changes involve substituting words with synonyms or related terms to neutralize adversarial prompts~\cite{SemanticSmooth,JailGuard}. For instance, SemanticSmooth~\cite{SemanticSmooth} converts all verbs into the past tense and substitutes both verbs \& nouns with their semantic equivalents.
\textit{Sentence-level} changes involve modifying and rewriting the entire input query to expose the embedded attack intent~\cite{NeMoGuardrails,SemanticSmooth,RigorLLM,RapidResponse,JailGuard,X-Guard}. For example, NeMo Guardrails~\cite{NeMoGuardrails} encodes the input prompt into a structured format to clarify the request's intent. RigorLLM~\cite{RigorLLM} paraphrases or summarizes the input to identify potential jailbreak attempts.
X-Guard~\cite{X-Guard} translates the non-English input into English.
Naturally, textual modifications can be applied simultaneously at multiple levels of granularity, including character, word, and sentence levels, such as in SemanticSmooth~\cite{SemanticSmooth} and JailGuard~\cite{JailGuard}.
From a broader perspective, dynamic guardrails can be viewed as a form of input/output augmentation that enhances the robustness of jailbreak detection by exposing hidden attack intents. 
}

\revise{
\subsection{Applicability}
The applicability of a guardrail is largely determined by its dependency on model internals. \textbf{White-box guardrails} require access to the target LLM’s internal states, such as gradients or hidden activations, to detect jailbreak attempts. For instance, GradSafe~\cite{GradSafe} analyzes gradient patterns of safety-critical parameters, while JBShield~\cite{JBShield} compares internal state divergences between benign and malicious queries. These methods can achieve high detection accuracy but are limited to scenarios where model internals are accessible. In contrast, \textbf{black-box guardrails} operate solely on input and output text, making them suitable for proprietary or API-based LLMs. Examples include Prompt Guard~\cite{PromptGuard}, which uses a lightweight classifier to filter malicious inputs, and WildGuard~\cite{WildGuard}, which moderates both prompts and responses via a separate LLM. Black-box approaches offer broader deployment flexibility but may sacrifice granularity in detection. This dimension is critical for practical deployment: organizations using closed-source models like GPT-4 must rely on black-box guardrails, while those with custom LLMs can leverage white-box methods for finer control. 
}

\revise{
\subsection{Explainability}
Explainability refers to the ability of a guardrail to provide transparent, interpretable rationales for its safety judgments. \textbf{Opaque guardrails}, such as many model-based classifiers, output binary decisions without justification, which can hinder trust and debugging. For example, MoJE~\cite{MoJE} employs simple linguistic statistical techniques to classify inputs as safe or unsafe without explaining its reasoning. While opaque guardrails can be efficient and effective, their lack of transparency may limit adoption in high-stakes applications where understanding failure modes is crucial.
In contrast, \textbf{explainable guardrails} enhance usability by providing reasoning traces or confidence scores. GuardReasoner~\cite{GuardReasoner} employs chain-of-thought reasoning to step through potential harms before rendering a judgment, while SelfDefend~\cite{SelfDefend} summarizes the intent of a query before assessing its risk. Such transparency helps developers understand failure modes, refine safety policies, and comply with regulatory requirements. 
}

%% file: sec/experiment.tex
\section{Benchmark \& Leaderboard}
\label{sec:experiment}

\begin{table}[t]
	\centering
	\caption{The details of our collected benchmark datasets.}
	\vspace{-1.0ex}
	\resizebox{0.85\columnwidth}{!}{
		\begin{tabular}{c|c|c}
			\hline \hline
			Dataset & \# Prompts & Jailbreak Methods \\
			\hline
			JailbreakHub \cite{DAN24} & 1000 & IJP \cite{DAN24} \\
			\hline
			\multirow{4}{*}{JailbreakBench~\cite{JailbreakBench}} & \multirow{4}{*}{100} & GCG~\cite{GCG23}, AutoDAN~\cite{AutoDAN24} \\ \cline{3-3}
			& & TAP~\cite{TAP23}, {LLM-Fuzzer}~\cite{LLM-Fuzzer24} \\ \cline{3-3}
			& & DrAttack~\cite{DrAttack24} \\ \cline{3-3}
            & & X-Teaming~\cite{X-Teaming25} \\ \hline
			MultiJail \cite{MultiJail23} & 315 & MultiJail \\ \hline
            SafeMTData~\cite{ActorAttack24} & 600 & ActorAttack~\cite{ActorAttack24} \\ \hline
			AlpacaEval \cite{alpaca} & 805 & Normal Prompts  \\ \hline
            OR-Bench \cite{OR-Bench} & 1000 & Normal Prompts \\ \hline \hline
		\end{tabular}
	}
	\label{tab:benchmark}
	\vspace{-3.0ex}
\end{table}

\begin{table*}[t]
	\centering
    \caption{The ASR ($\downarrow$) / PGR ($\downarrow$) results for the target LLM (Llama-3-8B-Instruct) with different guardrails against five major categories of jailbreak attacks, including row averages. (Pre) and (Post) denote the pre-processing and post-processing versions of the guardrails, respectively. (Direct) and (Intent) denote the direct prompt and intent prompt based versions of SelfDefend~\cite{SelfDefend}, respectively.
	}
	\vspace{-2.0ex}
	\resizebox{\textwidth}{!}{
		\begin{tabular}{l|c|cc|cc|cc|cc|c}
			\toprule
			\multirow{2}{*}{Guardrails} & Manual & \multicolumn{2}{c|}{Optimization-based} & \multicolumn{2}{c|}{Generation-based} &\multicolumn{2}{c|}{Implicit} &\multicolumn{2}{c|}{Multi-turn} & \multirow{2}{*}{Average}     \\ \cline{2-10}
			& IJP & GCG & AutoDAN & TAP & LLM-Fuzzer & DrAttack & MultiJail & ActorAttack & X-Teaming &   \\ \hline
            \revise{Llama-3-8B-Instruct}  & 0.078/- & 0.130/- & 0.020/- & 0.140/- & 0.490/- & 0.100/- & 0.044/- & 0.227/- & 0.910/- & 0.238/-  \\
            PerplexityFilter    & 0.078/1.000 & 0.100/0.620 & 0.020/1.000 & 0.140/1.000 & 0.480/1.000 & 0.100/1.000 & 0.044/1.000 & 0.227/1.000 & 0.960/1.000 & 0.239/0.958  \\
            SmoothLLM           & 0.115/0.261 & 0.020/0.020 & 0.030/0.110 & 0.140/0.170 & 0.500/0.810 & 0.150/0.660 & 0.032/0.575 & 0.893/0.893 & 0.850/0.910 & 0.303/0.490  \\
            Llama Guard (Pre)   & 0.062/0.563 & 0.100/0.390 & 0.020/0.480 & 0.140/0.460 & 0.450/0.680 & 0.100/0.840 & 0.044/0.952 & 0.220/0.967 & 0.910/1.000 & 0.227/0.704  \\
            Llama Guard (Post)  & 0.061/0.061 & 0.090/0.090 & 0.020/0.020 & 0.150/0.150 & 0.390/0.390 & 0.090/0.090 & 0.041/0.041 & 0.223/0.223 & 0.960/0.960 & 0.225/0.225  \\
            GradSafe            & 0.077/0.599 & 0.130/0.770 & 0.010/0.040 & 0.070/0.580 & 0.450/0.580 & 0.100/0.420 & 0.044/0.917 & 0.188/0.863 & 0.950/0.990 & 0.224/0.640  \\
            GradientCuff        & 0.016/0.058 & 0.080/0.140 & \textbf{0.000}/0.020 & 0.070/0.090 & 0.360/0.480 & 0.030/0.130 & \textbf{0.016}/0.149 & 0.118/0.648 & \textbf{0.640}/\textbf{0.710} & 0.148/0.269  \\
            SelfDefend (Direct) & 0.020/0.267 & 0.030/0.080 & 0.010/0.120 & 0.080/0.180 & 0.090/0.140 & 0.070/0.590 & 0.038/0.752 & 0.133/0.702 & 0.970/0.990 & 0.160/0.425  \\
            SelfDefend (Intent) & 0.022/0.285 & 0.030/0.070 & 0.010/0.130 & 0.120/0.180 & 0.030/0.130 & \textbf{0.010}/0.130 & 0.032/0.584 & 0.152/0.767 & 0.940/0.970 & 0.150/0.361  \\
            WildGuard (Pre)     & 0.004/0.033 & 0.020/0.020 & \textbf{0.000}/0.020 & \textbf{0.030}/0.060 & \textbf{0.000}/\textbf{0.000} & 0.080/0.500 & 0.044/0.797 & 0.150/0.757 & 0.960/0.980 & 0.143/0.352  \\
            WildGuard (Post)    & 0.020/0.020 & 0.050/0.050 & 0.010/0.010 & 0.060/0.060 & 0.090/0.090 & 0.060/0.060 & 0.035/0.035 & 0.148/0.148 & 0.950/0.950 & 0.158/0.158  \\
            Prompt Guard        & \textbf{0.000}/\textbf{0.000} & \textbf{0.000}/0.080 & 0.020/0.420 & 0.170/0.940 & \textbf{0.000}/\textbf{0.000} & 0.100/0.940 & 0.044/1.000 & 0.225/0.995 & 0.910/1.000 & 0.163/0.597  \\
            GuardReasoner (Pre) & \textbf{0.000}/0.009 & \textbf{0.000}/\textbf{0.000} & \textbf{0.000}/0.010 & \textbf{0.030}/0.070 & 0.010/0.020 & 0.080/0.360 & 0.029/0.349 & 0.143/0.740 & 0.920/0.960 & \textbf{0.135}/0.280  \\
            GuardReasoner (Post)& 0.023/0.023 & 0.040/0.040 & \textbf{0.000}/\textbf{0.000} & 0.050/\textbf{0.050} & 0.050/0.050 & 0.030/\textbf{0.030} & 0.022/\textbf{0.022} & \textbf{0.107}/\textbf{0.107} & 0.950/0.950 & 0.141/\textbf{0.141}  \\
			\bottomrule
		\end{tabular}
	}
	\label{tab:results}
	\vspace{-3.0ex}
\end{table*}

\subsection{Evaluation Setup}
\label{sec:eval_setup}

\noindent
\textbf{Datasets and Target Models.}
Based on the five categories of existing jailbreak attacks we surveyed in \S\ref{sec:jailbreak} — manual, optimization-based, generation-based, implicit, and multi-turn jailbreaks — we identify representative jailbreak attack methods in each category.
We then collect six benchmark datasets, \textbf{JailbreakHub}~\cite{DAN24}, \textbf{JailbreakBench}~\cite{JailbreakBench}, \textbf{SafeMTData}~\cite{ActorAttack24}, \textbf{MultiJail}~\cite{MultiJail23}, \textbf{AlpacaEval}~\cite{alpaca} and \textbf{OR-Bench}~\cite{OR-Bench}, from which we use their user prompts \revise{to build test queries for diverse guardrails.}
Table~\ref{tab:benchmark} lists the details of our collected benchmark datasets.
\textit{JailbreakHub} is a framework that collects and categorizes wild jailbreak prompts designed to bypass safety restrictions in LLMs. \revise{To refine ASR/PGR to three decimal places, we} randomly sample 1,000 in-the-wild prompts (IJP) from JailbreakHub as manual attacks.
\textit{JailbreakBench} is an open-source robustness benchmark specifically designed to evaluate and measure the vulnerability of LLMs to jailbreak attacks. We use \revise{all} 100 harmful instructions from JailbreakBench to drive optimization-based jailbreaks (GCG~\cite{GCG23} and AutoDAN~\cite{AutoDAN24}), generation-based jailbreaks (TAP~\cite{TAP23} and LLM-Fuzzer~\cite{LLM-Fuzzer24}), implicit jailbreak attack DrAttack~\cite{DrAttack24}, and multi-turn jailbreak attack X-Teaming~\cite{X-Teaming25}.
\textit{SafeMTData} is a dataset contains the initial multi-turn jailbreak prompts produced by ActorAttack~\cite{ActorAttack24}. We choose \revise{all} 600 queries from SafeMTData as multi-turn jailbreak attacks.
\textit{MultiJail} is the first manually-created multilingual jailbreak dataset, covering both high-resource and low-resource languages. We choose \revise{all} 315 jailbreak prompts in Bengali from MultiJail as multilingual jailbreaks.
\textit{AlpacaEval} is an automatic evaluation framework designed to assess the performance of instruction-following LLMs. We use \revise{all} 805 instructions from AlpacaEval as normal prompts.
\textit{OR-Bench} is the first large-scale benchmark measuring LLMs' over-refusal of 80,000 seemingly toxic but benign prompts across multiple categories. We randomly select 1,000 prompts from OR-Bench as normal prompts \revise{to also calculate ASR/PGR to three decimal places.}
\revise{We evaluate the defense performance of guardrails on open-source LLMs (i.e., \textbf{Llama-3-8B-Instruct}~\cite{llama3} and \textbf{Vicuna-13b-v1.5}~\cite{Vicuna}) and a closed-source LLM (i.e., \textbf{GPT-4-0125-Preview}~\cite{GPT4}), covering diverse model architectures and sizes.}
\revise{For evaluations on Vicuna-13b-v1.5 and GPT-4-0125-Preview, please refer to Appendix~\ref{sec:vicuna_results} and Appendix~\ref{sec:gpt4_results}, respectively.}

\noindent
\textbf{Metrics.}
We measure the defense performance of guardrails by the attack success rate (ASR), i.e., the frequency with which jailbreak prompts in a benchmark dataset bypass the guardrail and the target LLM. The lower the ASR, the stronger the defense performance.
Adopting the methodology from prior studies (\cite{qi2023fine,ActorAttack24,X-Teaming25}), we leverage GPT-4o's robust evaluation capabilities for assessment. This involves supplying GPT-4o, acting as a judge, with harmful objectives and the corresponding model outputs. The judge assigns a score from 1 to 5, where higher scores denote more harmful and detailed model responses. An attack is deemed successful if it receives a score of 5 from the GPT-4o Judge. For a detailed explanation of the scoring rubric, please see \cite{qi2023fine,ActorAttack24}.

\noindent
\textbf{Attack Configuration.}
To assess the jailbreak defense performance of guardrails, we employ the most widely used jailbreak attacks, including a manual attack (\textbf{IJP}~\cite{DAN24}), optimization-based attacks (\textbf{GCG}~\cite{GCG23} and \textbf{AutoDAN}~\cite{AutoDAN24}), generation-based attacks (\textbf{TAP}~\cite{TAP23} and \textbf{LLM-Fuzzer}~\cite{LLM-Fuzzer24}), implicit attacks (\textbf{DrAttack}~\cite{DrAttack24} and \textbf{MultiJail}~\cite{MultiJail23}), and multi-turn attacks (\textbf{ActorAttack}~\cite{ActorAttack24} and \textbf{X-Teaming}~\cite{X-Teaming25}).
In the context of \textit{IJP}, 1,000 adversarial queries were randomly sampled from the forbidden question set with jailbreak prompts~\cite{DANDataSet}, curated by JailbreakHub.
Regarding \textit{GCG}, its individual variant was selected, and the adversarial suffix was optimized against the target LLM employing a batch size of 512 and subjected to 500 optimization iterations.
For the \textit{AutoDAN} methodology, the hierarchically-guided genetic algorithm variant, specifically AutoDAN-HGA, was adopted. The genetic algorithm integral to AutoDAN-HGA operates with a crossover probability of 0.5, a mutation probability of 0.01, and undergoes 500 optimization iterations.
For GCG and AutoDAN, we migrate jailbreak prompts optimized for Vicuna-13b to attack GPT-4.
Concerning \textit{TAP}, the Vicuna-13b-v1.5 model~\cite{Vicuna} was utilized as the attacking agent. The parameters for TAP were configured with a maximum depth of 5, a maximum width of 5, and a branching factor of 4. The designated target models for TAP included Llama-3-8B-Instruct~\cite{llama3}, Vicuna-13b-v1.5~\cite{Vicuna} or GPT-4-0125-Preview~\cite{GPT4}.
In the case of \textit{LLM-Fuzzer}, GPT-3.5 served as the auxiliary model for generating mutational inputs, and the query limit directed at the target LLMs was established at 200.
For \textit{DrAttack}, jailbreak prompts were formulated using GPT-4o.
With respect to \textit{MultiJail}, the entirety of the 315 available queries in the Bengali language was selected.
For the \textit{ActorAttack} strategy, a corpus of 600 queries was sourced from the SafeMTData dataset~\cite{ActorAttack24} (specifically, the \texttt{SafeMTData/Attack\_600.json} file available on Hugging Face).
For \textit{X-Teaming}, we set the attacking model as Qwen2.5-32B-Instruct~\cite{Qwen2.5} and use the TextGrad-based text optimization to refine jailbreak prompts.
Regarding \textit{AlpacaEval}, all 805 questions within the AlpacaEval dataset were utilized.
For \textit{OR-Bench}, a subset of 1,000 prompts was randomly selected from the OR-Bench dataset~\cite{OR-Bench} (specifically, the \texttt{or-bench-80k.csv} file on Hugging Face).

It is pertinent to note that: \textit{The prompts associated with IJP, MultiJail, ActorAttack, AlpacaEval, and OR-Bench are static in nature. Consequently, all guardrail mechanisms encounter identical input stimuli, irrespective of whether they are safeguarding Llama-3-8B, Vicuna-13b or GPT-4. In contrast, GCG, AutoDAN, and DrAttack are specifically tailored to either Llama-3-8B, Vicuna-13b or GPT-4. As such, guardrails receive uniform inputs when defending the same designated target LLM. Conversely, TAP, LLM-Fuzzer, and X-Teaming represent adaptive attack methodologies. This implies that guardrail systems are presented with varied inputs, even when applied to the identical target LLM.}

\begin{figure*}[t]
	\centering
	\subfloat[Delay]{\label{fig:delay}
		\includegraphics[width=0.3\textwidth]{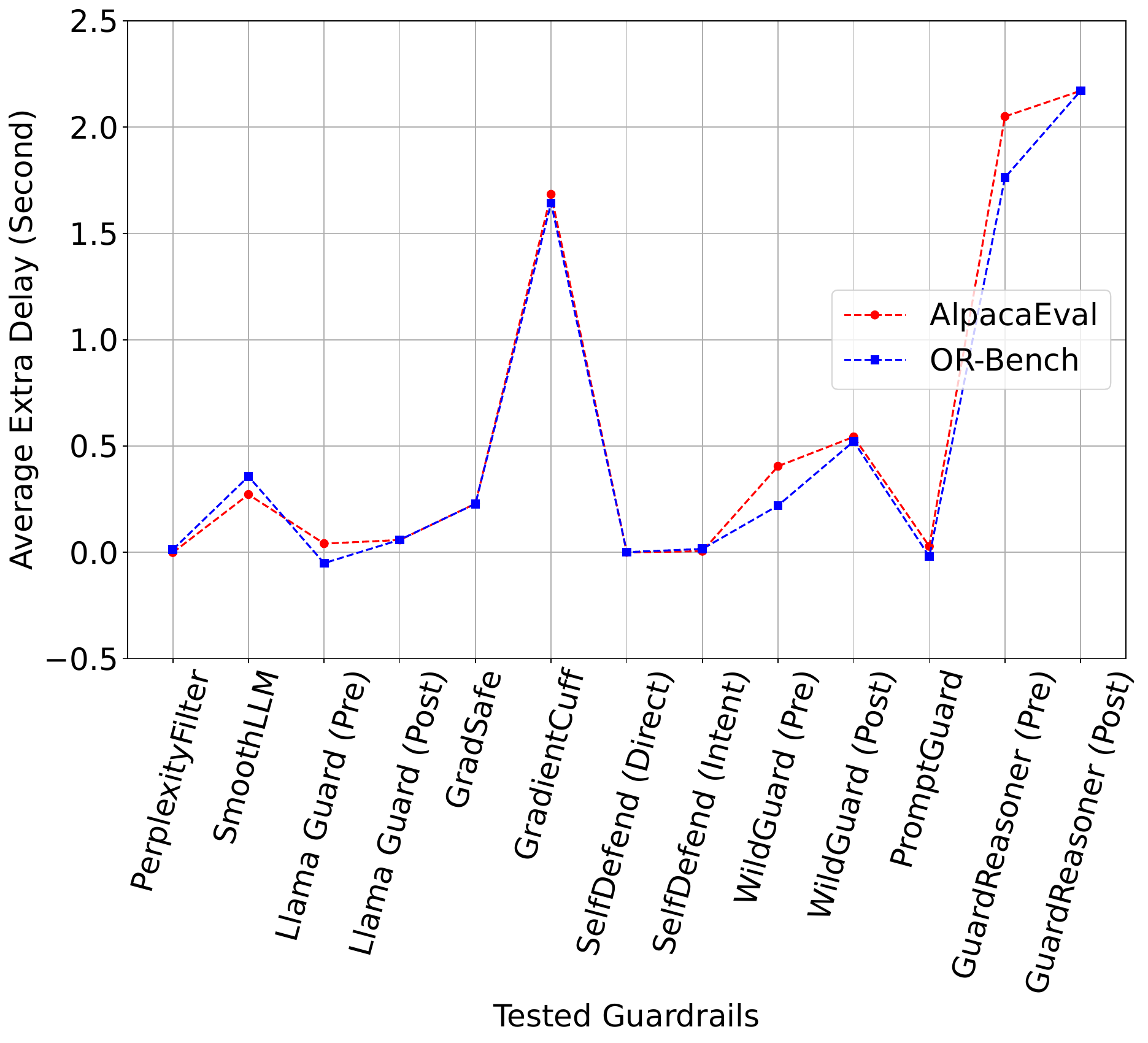}}
	\subfloat[Memory]{\label{fig:memory}
		\includegraphics[width=0.29\textwidth]{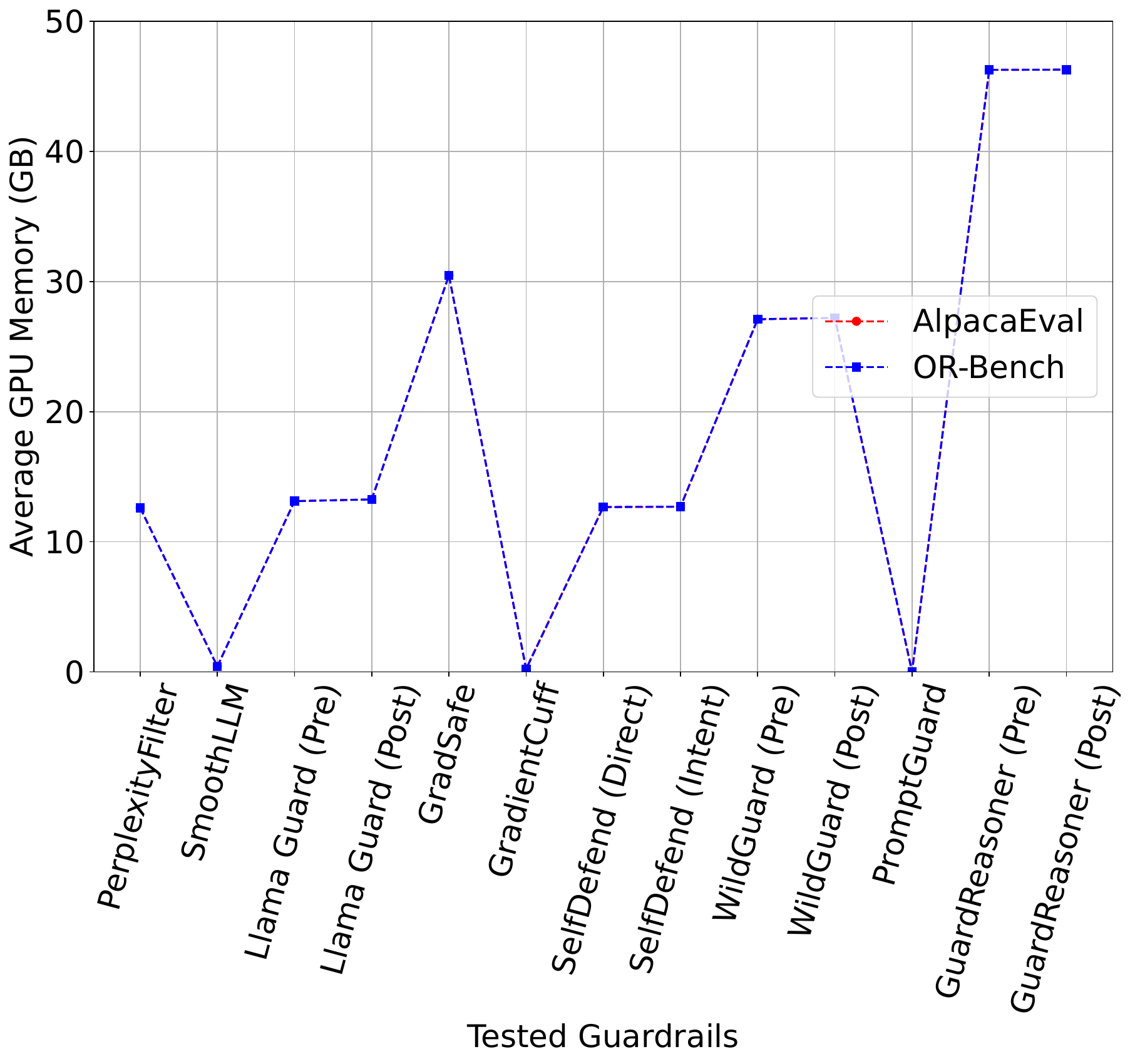}}
	\subfloat[Utility]{\label{fig:utility}
		\includegraphics[width=0.35\textwidth]{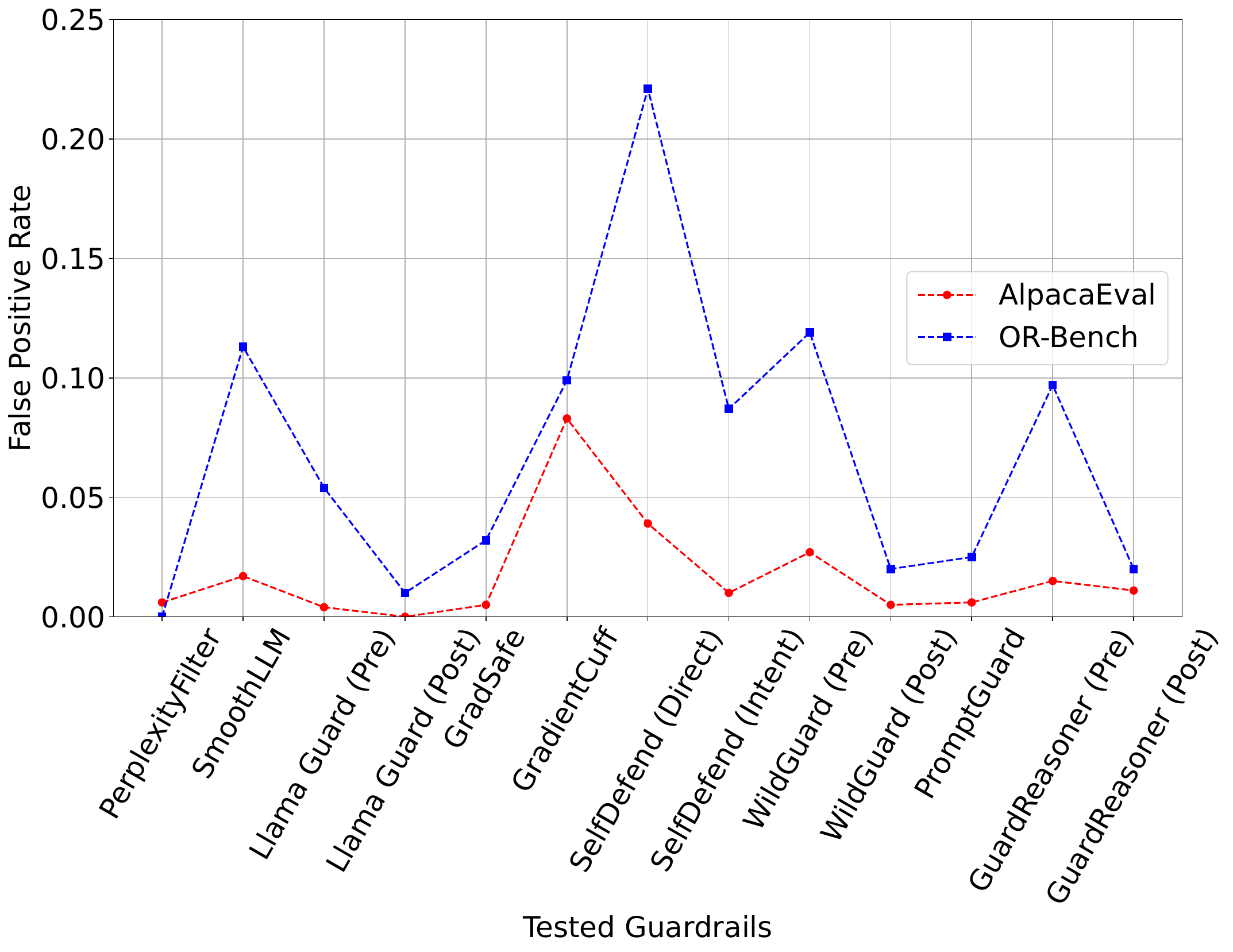}}
	\caption{The delay, memory usage, and utility of guardrails.}
	\label{fig:fig_delay_memory_utility}
	\vspace{-3.0ex}
\end{figure*}

\noindent
\textbf{Baselines.}
We \revise{evaluate} our framework with popular jailbreak defense methods, including \textbf{Perplexity Filter}~\cite{PerplexityFilter}, \textbf{SmoothLLM}~\cite{SmoothLLM}, \textbf{Llama Guard}~\cite{LlamaGuard}, \textbf{GradSafe}~\cite{GradSafe}, \textbf{GradientCuff}~\cite{GradientCuff}, \textbf{SelfDefend}~\cite{SelfDefend}, \textbf{WildGuard}~\cite{WildGuard}, \textbf{Prompt Guard}~\cite{PromptGuard}, and \textbf{GuardReasoner}~\cite{GuardReasoner}.
Specifically, \textit{Perplexity Filter} leverages a Llama-2-7b model to calculate the perplexity of the input prompt. A jailbreak is considered to happen when the perplexity exceeds a threshold. We set this threshold at the maximum perplexity of any prompt in the JailbreakBench dataset of harmful behavior prompts.
\textit{SmoothLLM} perturbs the jailbreak prompts with character-level changes to enable the target LLM to perform defense. In this paper, we set SmoothLLM to conduct character swapping with a 10\% perturbation percentage.
\textit{Llama Guard} is a fine-tuned Llama-2-7b model designed to detect the toxicity category of input prompts.
\textit{GradSafe} is a gradient-based detection method that identifies unsafe or jailbreak prompts in LLMs by analyzing the consistent gradient patterns of safety-critical parameters when paired with compliance responses.
\textit{GradientCuff} is a method for detecting jailbreak attacks on LLMs by analyzing the refusal loss landscape, leveraging gradient-based patterns to identify and block adversarial prompts while maintaining normal query performance.
\textit{SelfDefend} is a practical jailbreak defense framework for LLMs that uses a shadow LLM instance to concurrently detect harmful queries while the target LLM processes them, providing robust protection with minimal delay.
\textit{WildGuard} is an open, lightweight, multi-task moderation tool for LLMs that detects malicious user prompts, harmful model responses, and model refusal behavior.
\textit{Prompt Guard} is a security tool developed by Meta that detects and blocks malicious inputs (e.g., jailbreak attempts, prompt injections) in LLM applications, using a lightweight classifier model Prompt-Guard-86M to filter harmful content in real time.
\textit{GuardReasoner} is a reasoning-based guard model designed to enhance the safety of LLMs by integrating explicit step-by-step reasoning into the moderation process.
Our evaluations are implemented using PyTorch 2.6.0 and conducted on NVIDIA Hopper H800 GPUs.





\subsection{Benchmark Evaluation}
\label{sec:benchmark_eval}

\noindent
\textbf{Defense Performance.}
\label{sec:benchmark_defense}
We first analyze the defense performance of various guardrails. As delineated in Table~\ref{tab:results}, which presents the ASR, a lower value indicates superior defense capabilities. On average, {GuardReasoner (Pre)} demonstrates the most robust defense, achieving the lowest ASR of 0.135. Following closely is {GuardReasoner (Post)}, underscoring the efficacy of the reasoning process prior to safety determination inherent in the GuardReasoner framework. Conversely, {SmoothLLM} exhibits the highest ASR of 0.303, rendering it the least effective in this cohort. This suboptimal performance may be attributed to its mechanism of token-level input perturbation, which appears to be primarily effective against jailbreak techniques characterized by adversarial suffixes, such as GCG, while offering limited protection against a broader spectrum of attacks.

Shifting focus to PGR, presented in Table~\ref{tab:results}, {GuardReasoner (Post)} achieves the best PGR of 0.141. Despite its superior precision in identifying malicious inputs, {GuardReasoner (Post)} does not attain state-of-the-art (SOTA) overall defense performance. A plausible explanation is its potential operational overlap with the target LLM's intrinsic safety mechanisms. That is, there might be a significant number of instances where {GuardReasoner (Post)} identifies a response as safe, and concurrently, the target LLM also recognizes the harmful nature of the query and refuses to respond. \revise{Therefore, this overlap diminishes} the unique contribution of {GuardReasoner (Post)} to the ASR reduction when compared to a guardrail like {GuardReasoner (Pre)} which operates on a different paradigm.

\noindent
\textbf{Efficiency.}
\label{sec:benchmark_efficiency}
The efficiency of guardrails is a critical factor for practical deployment, which we assess in terms of latency and GPU memory consumption. Figure~\ref{fig:fig_delay_memory_utility}(a) illustrates the extra delay introduced by different guardrail methodologies when processing normal inputs from the AlpacaEval and OR-Bench datasets. {Perplexity Filter, Llama Guard, SelfDefend and PromptGuard} stand out with negligible latency. In contrast, {GuardReasoner} and {GradientCuff} impose the most significant delays, with {GuardReasoner (Post)} being particularly notable. This suggests that the profound reasoning capabilities that afford {GuardReasoner} its enhanced defense performance come at the cost of increased processing time. The majority of other guardrails maintain an additional delay generally not exceeding 0.5 seconds.

From the perspective of GPU memory utilization, depicted in Figure~\ref{fig:fig_delay_memory_utility}(b), {GuardReasoner} again registers the highest memory footprint, consistent with its complex reasoning architecture. Conversely, {SmoothLLM}, {GradientCuff}, and {PromptGuard} are the most memory-efficient, with their consumption approaching negligible levels. This highlights a clear trade-off between the sophistication of the defense mechanism and its resource intensiveness.

\noindent
\textbf{Utility.}
\label{sec:benchmark_utility}
Beyond security and efficiency, the utility of a guardrail, specifically its ability not to impede benign user interactions, is paramount. We measure the FPR on AlpacaEval and OR-Bench datasets, as shown in Figure~\ref{fig:fig_delay_memory_utility}(c). A higher FPR indicates a greater propensity to incorrectly flag legitimate prompts as malicious. {SelfDefend (Direct)} exhibits the highest FPR on OR-Bench, at 0.221. On AlpacaEval, {GradientCuff} records the highest FPR of 0.083. These figures suggest that these guardrails have a higher likelihood of intercepting normal user queries. Other guardrails with comparatively high FPRs include {SmoothLLM}, {SelfDefend (Intent)}, {WildGuard (Pre)}, and {GuardReasoner (Pre)}. Although these three methods demonstrate strong defense performance (low ASR), their elevated FPRs underscore a critical trade-off between security and utility. Systems that are highly stringent in blocking threats may inadvertently penalize legitimate interactions, diminishing the overall user experience.

\begin{figure}[t]
	\centering
	\includegraphics[width=0.9\columnwidth]{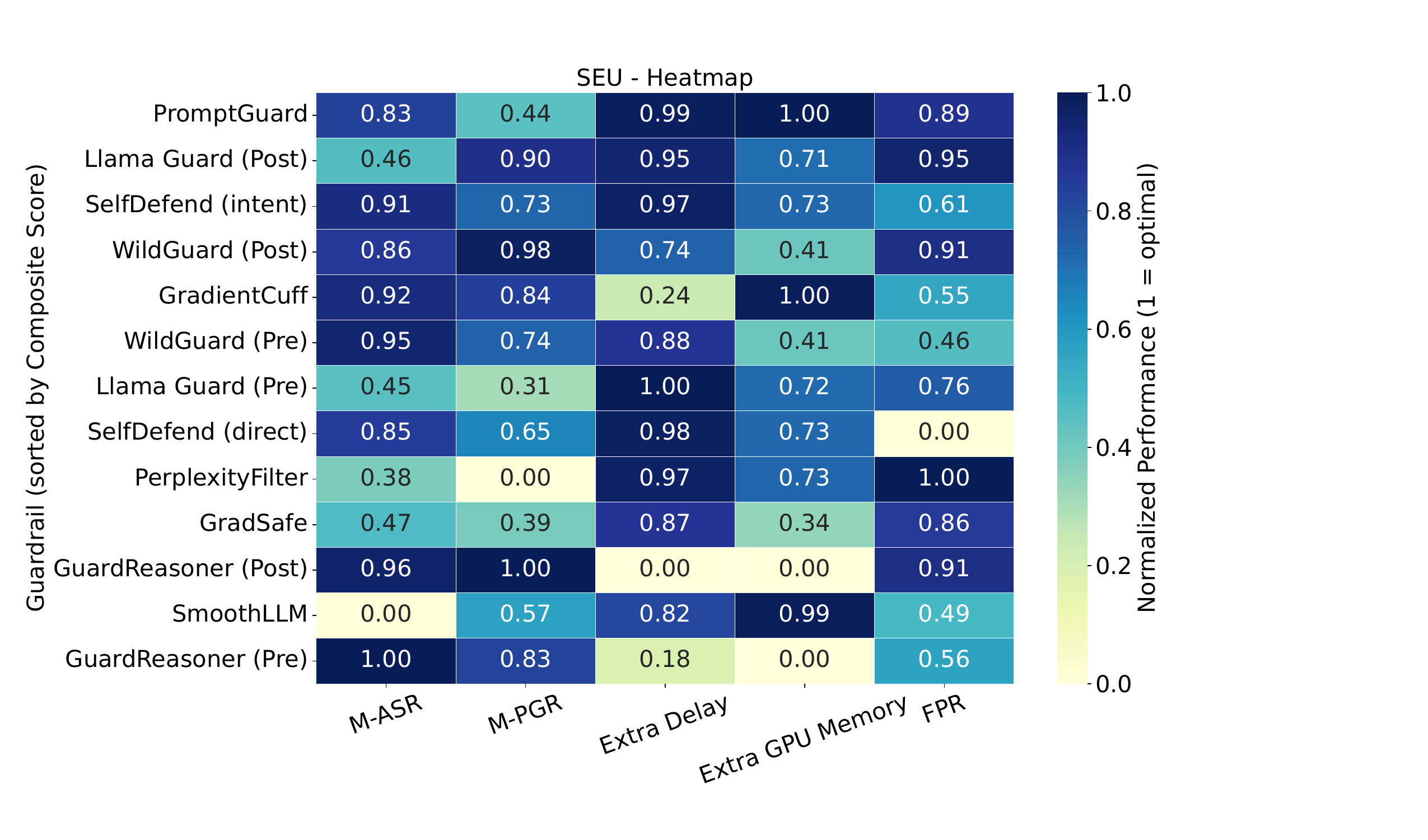}
	\caption{The heatmap of guardrails.}
	\label{fig:heatmap}
\end{figure}

\subsection{Leaderboard on SEU}
\label{sec:leaderboard}
To provide a holistic evaluation, we compare guardrails across five key metrics: ASR, PGR, Extra Delay, GPU Memory, and FPR. We average ASR and PGR over the nine jailbreak attacks (cf. Table~\ref{tab:results}) to derive Mean-ASR (M-ASR) and Mean-PGR (M-PGR). The other metrics are measured using the OR-Bench dataset.
For a unified ranking, we normalize each metric to a [0, 1] range and invert the scores (1 - normalized value), ensuring higher values consistently indicate better performance. We then compute a {Composite Score} for each guardrail by averaging these five transformed scores. This score underpins the ranking visualized in the heatmap in Figure~\ref{fig:heatmap}.

The analysis reveals inherent trade-offs, as no single guardrail excels across
all dimensions. For instance, {PromptGuard} achieves the highest Composite Score
but its low M-PGR suggests potential gaps in detection robustness. Conversely,
{GuardReasoner (Pre)} ranks lower but provides superior defense (high M-ASR and
M-PGR) at a significant cost to efficiency and utility. {SelfDefend (Intent)}
offers a balanced profile, with its main weakness being a higher FPR. This
leaderboard underscores that the optimal guardrail choice is context-dependent,
contingent on the specific security requirements and operational constraints of
a given deployment scenario. We believe this leaderboard will serve as a
valuable resource for practitioners in selecting appropriate guardrails based on
their unique needs.

\section{Practical Insights \& Implications}
\label{sec:practical_implications}

\noindent
\textbf{Answer to RQ3: Session-level Guardrails v.s. Multi-turn Jailbreaks.}
\label{sec:session_guardrail}
A critical question arises regarding session-level guardrails: given their reliance on LLM dialogue history (both input and output) for threat assessment, how effectively do they counter sophisticated multi-turn jailbreak attacks?
Our analysis, focusing on three session-level guardrails—{Llama Guard (Post)}, {WildGuard (Post)}, and {GuardReasoner (Post)}—reveals nuanced performance. As indicated in Table~\ref{tab:results}, these guardrails maintain an ASR above 10\% against the {ActorAttack} multi-turn jailbreak. 
Furthermore, when faced with the more adaptive {X-Teaming} attack, the ASR for most guardrails, including these session-level ones, exceeds 90\%. {GradientCuff} is a partial exception with a 64\% ASR, but this is still a high failure rate. 
These findings underscore a significant vulnerability of current session-level guardrails against advanced multi-turn attacks. The high ASR, particularly against adaptive attacks like {X-Teaming}, suggests that these defenses can be readily bypassed if the attack unfolds over several interactions. This highlights an urgent imperative to develop more robust guardrail methodologies specifically designed to address the evolving landscape of multi-turn jailbreaks.

\begin{figure}[t]
	\centering
	\includegraphics[width=0.8\columnwidth]{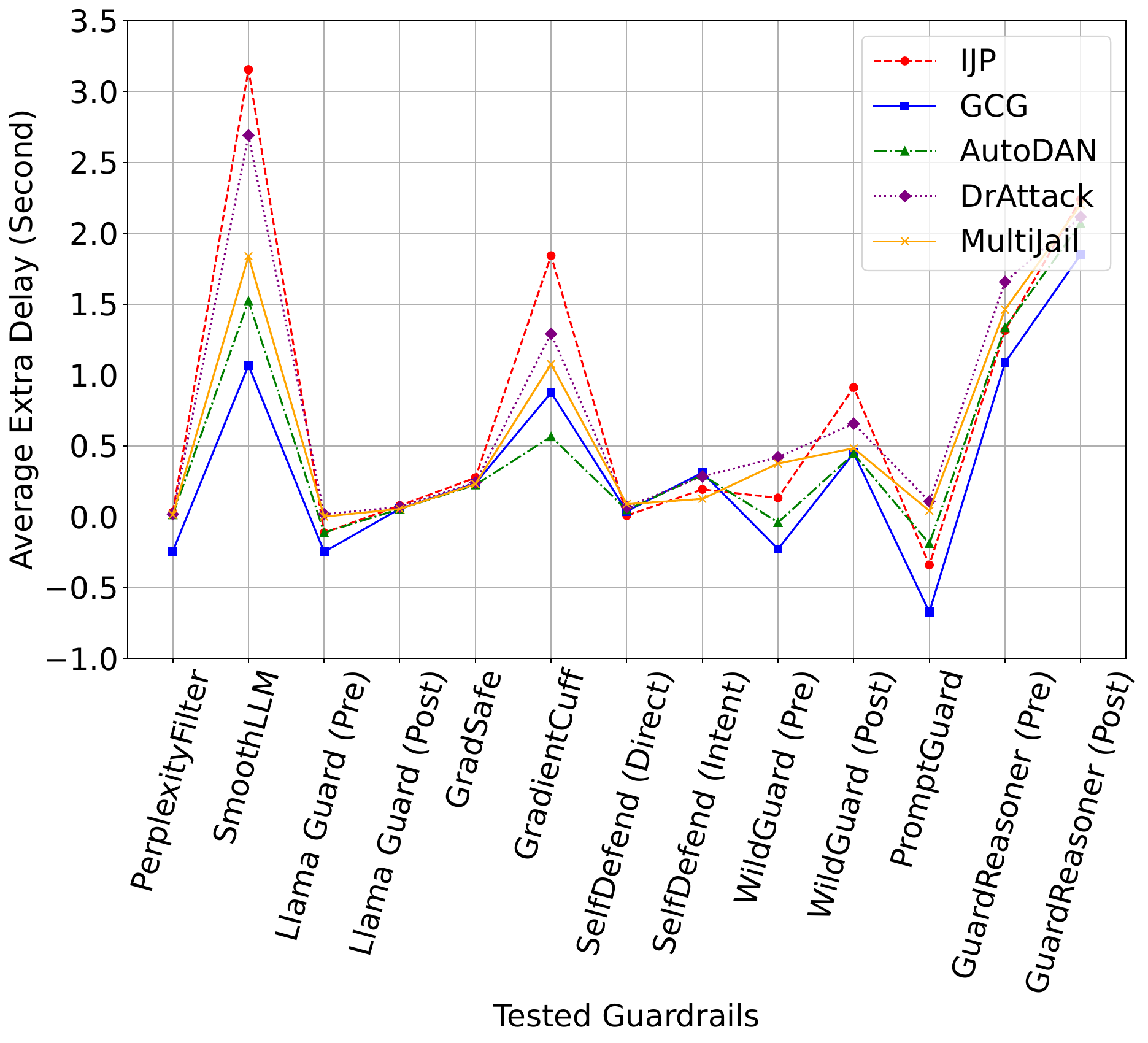}
	\vspace{-1.0ex}
	\caption{The delay of guardrails against different attack types.}
	\label{fig:delay_attack}
\end{figure}

\noindent
\textbf{Answer to RQ1: Intervention Stages on Delay.}
\label{sec:intervention_delay}
The intervention stage of a guardrail—whether it operates pre-processing (on user input), intra-processing (during LLM generation), or post-processing (on LLM output)—can significantly impact system latency. We investigate this relationship by examining the data presented in Figure~\ref{fig:delay_attack}.
Observations indicate that, with the notable exception of {GuardReasoner (Pre)}, pre-processing guardrails such as {Perplexity Filter}, {Llama Guard (Pre)}, {SelfDefend (Direct)}, {SelfDefend (Intent)}, {WildGuard (Pre)}, and {Prompt Guard} generally introduce negligible, or in some cases even negative, additional latency. The higher latency of {GuardReasoner (Pre)} is attributable to its more complex reasoning processes. In contrast, intra-processing and post-processing guardrails exhibit more varied latency profiles relative to each other.
A key finding is that for identical detection models, post-processing variants consistently incur greater delay than their pre-processing counterparts (e.g., {WildGuard (Post)}'s delay is greater than that of {WildGuard (Pre)}). This phenomenon arises because post-processing methods inherently must await the completion of the target LLM's generation phase before they can intervene. Conversely, pre-processing guardrails possess the advantage of potentially halting the LLM's generation process immediately upon detecting a malicious input, thereby conserving computational time. Consequently, pre-processing guardrails, particularly those not reliant on extensive reasoning, generally offer a more latency-efficient solution for integrating safety measures.

\noindent
\textbf{Answer to RQ2: Technical Paradigms on GPU Memory Usage.}
\label{sec:technical_memory}
The underlying technical paradigm of a guardrail—be it rule-based, traditional model-based, or LLM-based—is expected to influence its GPU memory footprint. We examine this correlation using data from Figure~\ref{fig:fig_delay_memory_utility}.
The results show that the rule-based {SmoothLLM} incurs zero additional memory overhead, representing the most memory-efficient approach. Certain traditional model-based methods, specifically {GradientCuff} and {PromptGuard}, also demonstrate near-zero memory consumption, highlighting their lightweight nature. However, the landscape for model-based approaches is not uniform; {GradSafe}, another model-based technique, exhibits higher memory usage than several LLM-based methods, indicating significant variability in resource demands even within this category.
As anticipated, LLM-based guardrails generally impose a greater memory burden. This is an intrinsic consequence of their design, which necessitates loading and executing a large language model for safety inference. This observation aligns with the expectation that leveraging large language models for safety assessment incurs a higher resource cost in terms of memory. While rule-based and optimized model-based solutions offer substantial memory efficiency, the choice of paradigm must be carefully weighed against the desired detection capabilities and specific deployment constraints.

\noindent
\textbf{Answer to RQ4: Safety Granularity on Utility.}
\label{sec:safety_granularity}
The granularity at which a guardrail performs its safety checks—whether at the token-level, sequence-level (assessing the entire input or output), or session-level (considering the dialogue history)—may significantly affect its utility, particularly its propensity to misclassify benign prompts, as measured by the FPR. This aspect is explored using data from Figure~\ref{fig:fig_delay_memory_utility}(c).
Token-level guardrails, exemplified by {SmoothLLM} (which analyzes keywords in LLM responses) and {SelfDefend (Direct)} (which inspects harmful segments within queries), demonstrate relatively pronounced FPRs. Notably, {SelfDefend (Direct)} records the highest FPR on the {OR-Bench} dataset, exceeding 20\%. This suggests that token-level mechanisms, while focused, may inadvertently penalize legitimate interactions due to a potential lack of broader contextual understanding.
A comparative analysis further reveals that for the same underlying detection model, session-level guardrails (typically denoted by a ``(Post)" suffix, leveraging both LLM input and output) consistently achieve markedly lower FPRs than their sequence-level counterparts (often denoted by a ``(Pre)" suffix, relying solely on input). For instance, {WildGuard (Pre)} exhibits an FPR above 10\% on {OR-Bench}, whereas the FPR for {WildGuard (Post)} remains below 5\%. While sequence-level guardrails display a wider range of FPRs—some high, some low—session-level approaches generally maintain low FPR values across the board.
These observations collectively suggest that session-level guardrails tend to offer superior utility by minimizing false positives. This improved performance is likely attributable to their comprehensive use of contextual information derived from the entire interaction history, enabling a more nuanced distinction between genuinely harmful prompts and benign ones that might share superficial characteristics with attacks.

\begin{figure}[!htbp]
	\centering
	\includegraphics[width=0.85\columnwidth]{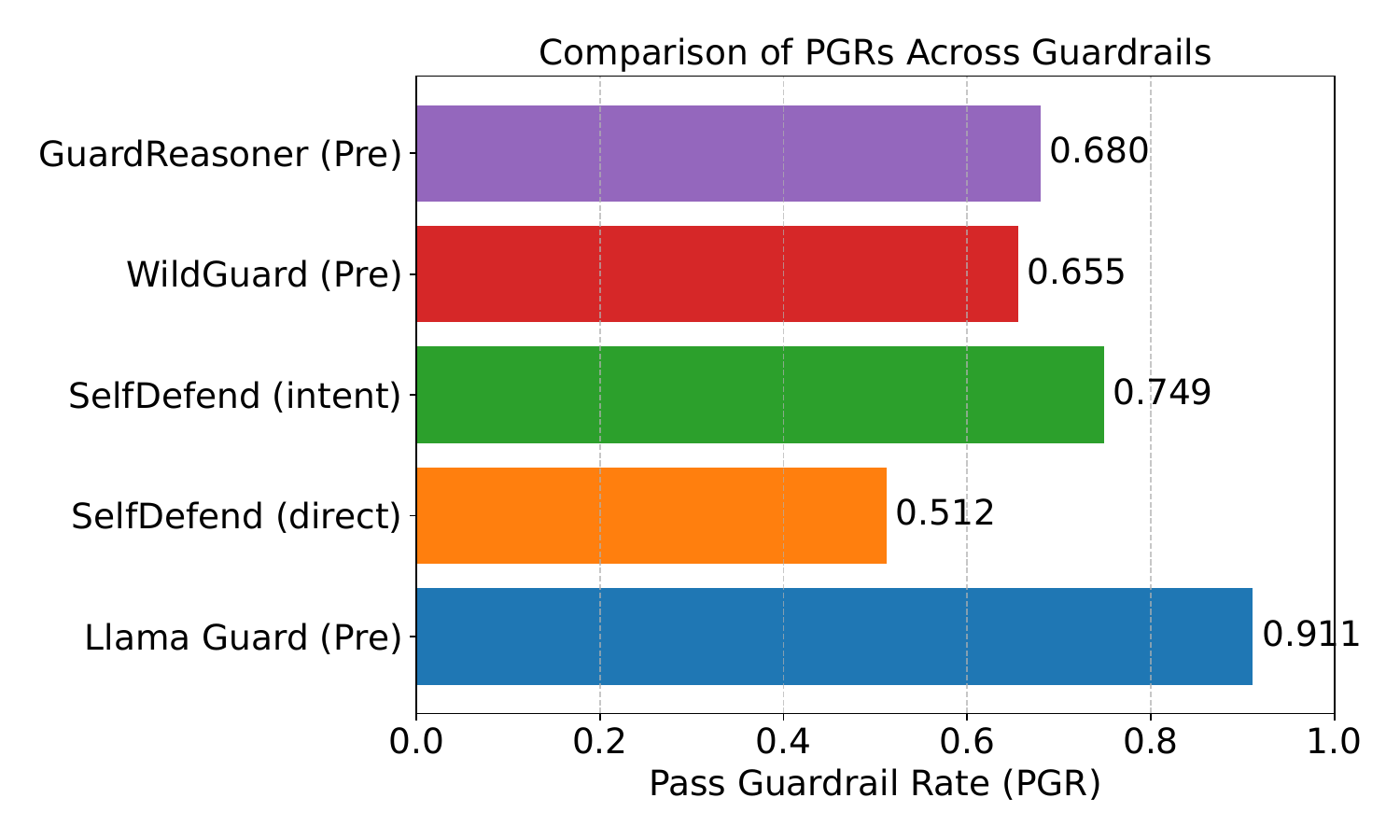}
	\caption{Cross-attack evaluation: injection attack on guardrails.}
	\label{fig:injection}
	\vspace{-1.0ex}
\end{figure}

\noindent
\textbf{Generalization: Cross-Attack Assessment.}
\label{sec:cross_attack_generalization}
While LLM-based guardrails have demonstrated efficacy against jailbreak attacks, a critical and often overlooked consideration is their robustness against other adversarial manipulations, specifically prompt injection attacks. Given that these guardrails are themselves powered by LLMs, their susceptibility to injection attacks—which could potentially subvert their safety assessment capabilities—presents a significant security concern. To investigate this, we evaluated the performance of LLM-based guardrails against 203 distinct injection attack samples sourced from the ``deepset/prompt-injections'' dataset on Hugging Face.

Our primary finding is that these injection attacks did not compromise the
fundamental operational integrity of the guardrails. That is, the guardrails
were not coerced into abandoning their safety analysis function to produce
arbitrary, irrelevant outputs (e.g., ``hello world''). They continued to process
the inputs for security threats as designed. However, their effectiveness in
identifying and mitigating these injections was limited. We measured the Pass
Guardrail Rate (PGR) for these attacks, with results presented in
Figure~\ref{fig:injection}. The data reveals that while LLM-based guardrails
exhibit a non-trivial capacity to filter prompt injections, this capability is
modest at best. This assessment underscores a crucial gap in the current state
of guardrail technology: the need for broader cross-attack generalization. For a
guardrail to be truly effective in practice, its defensive perimeter must extend
beyond jailbreak attempts to also detect and neutralize other forms of attacks
that could exploit its defense, such as prompt injections. This calls for the
development of more versatile and robust guardrail mechanisms capable of
addressing a wider spectrum of adversarial inputs.

%% file: sec/discussion.tex
\revise{
\section{Discussion}

\subsection{Limitations}
While this work provides a comprehensive analysis of LLM jailbreak guardrails, several limitations should be acknowledged. First, our efficiency evaluations were conducted on specific hardware configurations (NVIDIA H800 GPUs), and the absolute latency and memory consumption metrics may vary across different hardware platforms and optimization techniques such as quantization. However, the relative performance trends and trade-offs among guardrails are expected to remain consistent.

Besides, our evaluation relies on LLM-as-a-judge (GPT-4o) for assessing jailbreak success, which, while efficient for large-scale evaluation, may have inherent limitations in reliability. To validate this approach, we hired three Ph.D. students to evaluate the jailbreak success of 100 randomly selected attack samples. Their average agreement with the LLM's judgments was 0.9533, and their Fleiss' Kappa was 0.9319, indicating strong alignment. Nevertheless, future work could benefit from more robust evaluation methodologies that mitigate potential biases in automated assessment.



\subsection{Future Work}
Based on our analysis, we identify several promising directions for future research in LLM jailbreak guardrails:
\begin{itemize}[leftmargin=*]
    \item \textbf{Cross-Attack Robust Guardrails:} Current guardrails demonstrate limited generalization across different attack modalities. Future work should develop unified defense mechanisms that can effectively handle diverse threats, including jailbreaks, prompt injections, and other adversarial manipulations simultaneously. This requires guardrails that can recognize attack patterns beyond jailbreaks.
    \item \textbf{Adaptive and Self-Evolving Defenses:} As jailbreak techniques continuously evolve, fixed guardrails face the risk of rapid obsolescence. Research is needed on self-adaptive guardrails that can learn from new attack patterns and update their detection capabilities without complete retraining. This could include online learning approaches, anomaly detection in deployment environments, and automated defense refinement.
    \item \textbf{Multimodal and Cross-Modal Guardrails:} With the rise of multimodal LLMs, new vulnerabilities emerge at the intersection of different modalities. Future research should address the unique challenges of multimodal jailbreaks~\cite{JailBreakV} and develop guardrails that can analyze and protect across text, image, audio, and video inputs while maintaining efficiency.
\end{itemize}




}

%% file: sec/appendix.tex
\appendices

\begin{table*}[htbp]
	\centering
    \caption{The ASR ($\downarrow$) / PGR ($\downarrow$) results for the target LLM (Vicuna-13b-v1.5) with different guardrails against five major categories of jailbreak attacks, including row averages. (Pre) and (Post) denote the pre-processing and post-processing versions of the guardrails, respectively. (Direct) and (Intent) denote the direct prompt and intent prompt based versions of SelfDefend~\cite{SelfDefend}, respectively.
	}
	\vspace{-2.0ex}
	\resizebox{\textwidth}{!}{
		\begin{tabular}{l|c|cc|cc|cc|cc|c}
			\toprule
			\multirow{2}{*}{Guardrails} & Manual & \multicolumn{2}{c|}{Optimization-based} & \multicolumn{2}{c|}{Generation-based} &\multicolumn{2}{c|}{Implicit} &\multicolumn{2}{c|}{Multi-turn} & \multirow{2}{*}{Average}     \\ \cline{2-10}
			& IJP & GCG & AutoDAN & TAP & LLM-Fuzzer & DrAttack & MultiJail & ActorAttack & X-Teaming &   \\ \hline
            Vicuna-13b-v1.5     & 0.474/- & 0.890/- & 0.660/- & 0.530/- & 0.820/- & 0.780/- & 0.254/- & 0.238/- & 0.960/- & 0.649/-  \\
            PerplexityFilter    & 0.474/1.000 & 0.030/0.040 & 0.660/1.000 & 0.830/1.000 & 0.870/1.000 & 0.780/1.000 & 0.254/1.000 & 0.238/1.000 & 0.990/1.000 & 0.570/0.893 \\
            SmoothLLM           & 0.402/0.794 & 0.140/0.270 & 0.520/0.970 & 0.840/0.860 & 0.510/1.000 & 0.410/0.970 & 0.152/0.933 & 0.877/0.877 & 0.980/0.970 & 0.537/0.849 \\
            Llama Guard (Pre)   & 0.194/0.563 & 0.370/0.390 & 0.460/0.750 & 0.630/0.750 & 0.810/1.000 & 0.650/0.850 & 0.251/0.952 & 0.222/0.967 & 0.970/1.000 & 0.506/0.802 \\
            Llama Guard (Post)  & 0.250/0.250 & 0.400/0.400 & 0.610/0.610 & 0.600/0.600 & 0.830/0.830 & 0.390/0.390 & 0.248/0.248 & 0.230/0.230 & 0.970/0.970 & 0.503/0.503 \\
            GradSafe            & 0.471/0.994 & 0.890/1.000 & 0.660/1.000 & 0.580/0.960 & 0.900/1.000 & 0.780/1.000 & 0.254/1.000 & 0.238/1.000 & 0.980/1.000 & 0.639/0.995 \\
            GradientCuff        & 0.193/0.351 & 0.090/0.090 & 0.310/0.480 & 0.550/0.630 & 0.780/1.000 & 0.660/0.830 & \textbf{0.000}/\textbf{0.000} & 0.183/0.805 & 0.930/0.960 & 0.411/0.572 \\
            SelfDefend (Direct) & 0.050/0.262 & 0.080/0.080 & 0.020/0.080 & 0.210/0.270 & 0.190/0.270 & 0.330/0.480 & 0.187/0.743 & 0.132/0.720 & 0.960/0.990 & 0.240/0.433 \\
            SelfDefend (Intent) & 0.057/0.286 & 0.080/0.080 & 0.050/0.110 & 0.140/0.200 & 0.210/0.250 & \textbf{0.010}/0.090 & 0.127/0.568 & 0.157/0.763 & 0.960/1.000 & 0.199/0.372 \\
            WildGuard (Pre)     & 0.007/0.033 & 0.010/0.010 & \textbf{0.010}/0.020 & \textbf{0.040}/0.090 & \textbf{0.010}/\textbf{0.020} & 0.330/0.490 & 0.187/0.797 & 0.147/0.757 & 0.920/0.950 & 0.185/0.352 \\
            WildGuard (Post)    & 0.066/0.066 & 0.040/0.040 & 0.030/0.030 & 0.100/0.100 & 0.410/0.410 & 0.090/0.090 & 0.194/0.194 & 0.165/0.165 & 0.930/0.930 & 0.225/0.225 \\
            Prompt Guard        & \textbf{0.000}/\textbf{0.000} & 0.020/0.020 & 0.240/0.370 & 0.570/0.960 & 0.020/0.030 & 0.770/0.990 & 0.254/1.000 & 0.235/0.995 & 0.980/1.000 & 0.343/0.596 \\
            GuardReasoner (Pre) & \textbf{0.000}/0.009 & \textbf{0.000}/\textbf{0.000} & 0.020/0.020 & 0.050/0.080 & 0.040/0.040 & 0.150/0.270 & 0.057/0.349 & 0.143/0.740 & 0.940/0.960 & \textbf{0.156}/0.274 \\
            GuardReasoner (Post)& 0.050/0.050 & 0.030/0.030 & \textbf{0.010}/\textbf{0.010} & 0.060/\textbf{0.060} & 0.480/0.480 & 0.040/\textbf{0.040} & 0.060/0.060 & \textbf{0.100}/\textbf{0.100} & \textbf{0.900}/\textbf{0.900} & 0.192/\textbf{0.192} \\
			\bottomrule
		\end{tabular}
	}
	\label{tab:vicuna_results}
	\vspace{-1.0ex}
\end{table*}

\begin{figure}
	\centering
	\includegraphics[width=0.8\columnwidth]{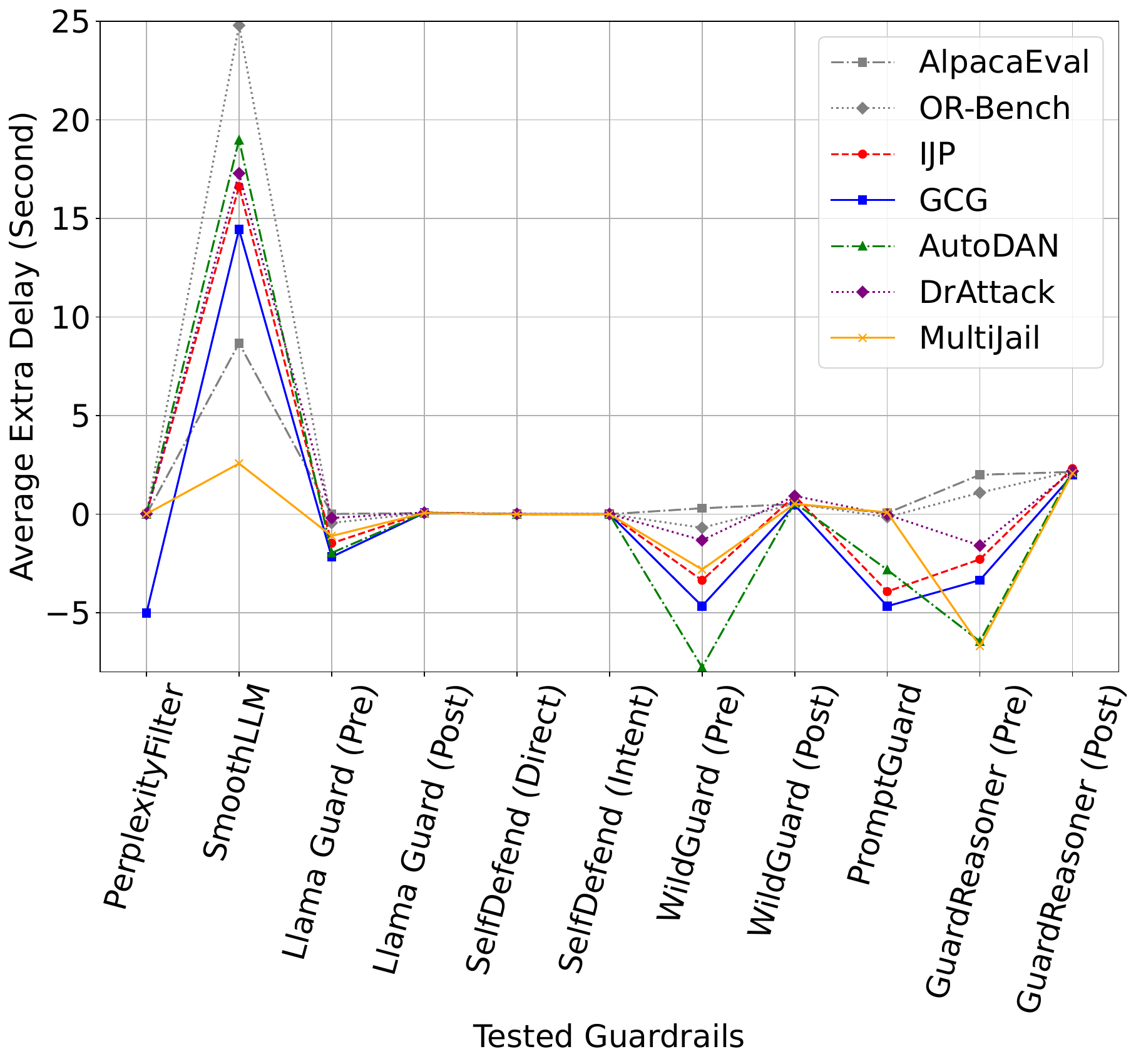}
	\caption{\revise{The delay of guardrails on GPT-4-0125-Preview. AlpacaEval and OR-Bench are two normal prompt datasets and others are jailbreak attacks.}}
	\label{fig:fig_delay_gpt}
\end{figure}

\begin{figure}
	\centering
	\includegraphics[width=0.798\columnwidth]{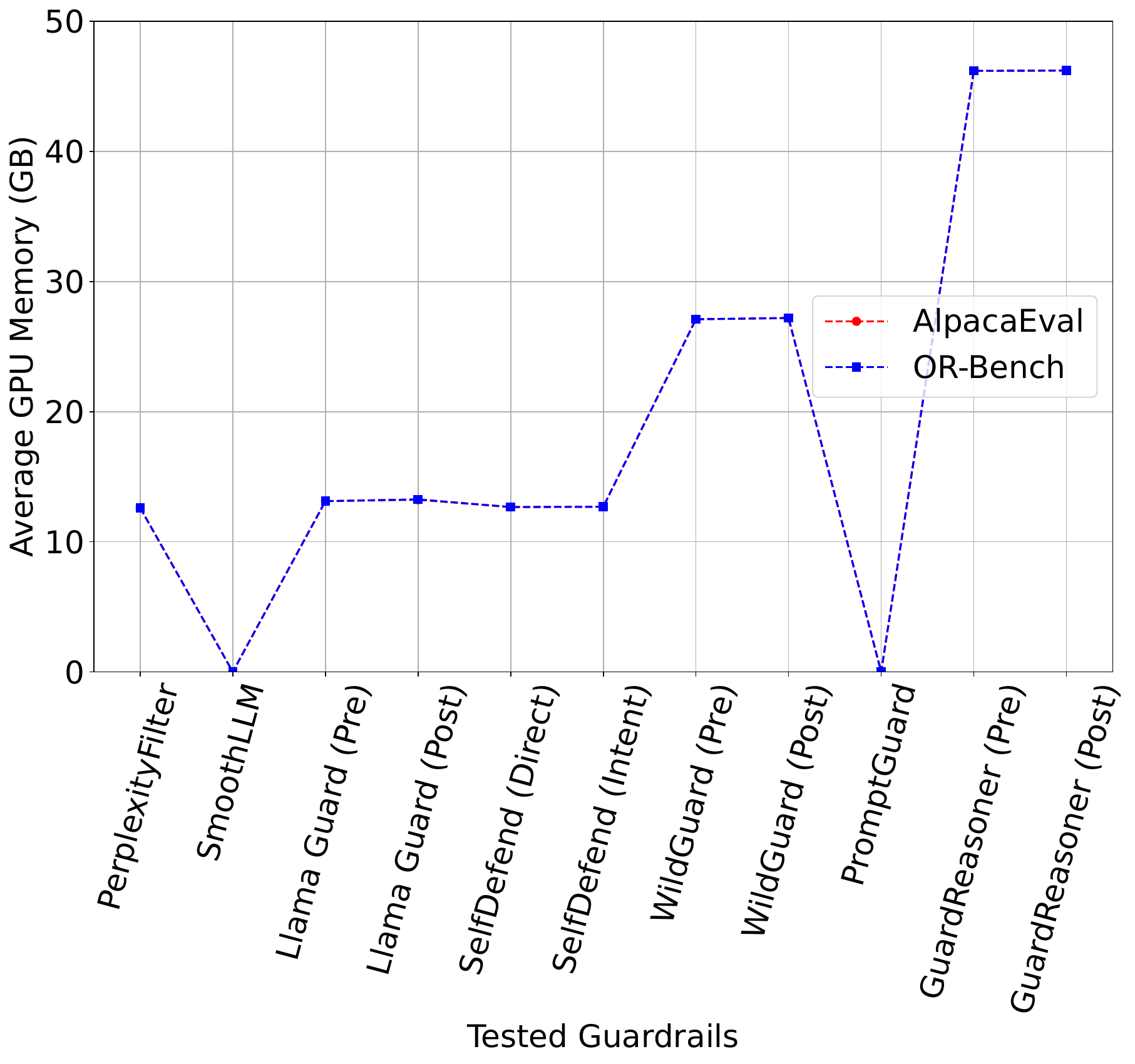}
	\caption{\revise{The memory usage of guardrails on GPT-4-0125-Preview.}}
	\label{fig:fig_memory_gpt}
\end{figure}

\begin{figure}
	\centering
	\includegraphics[width=0.82\columnwidth]{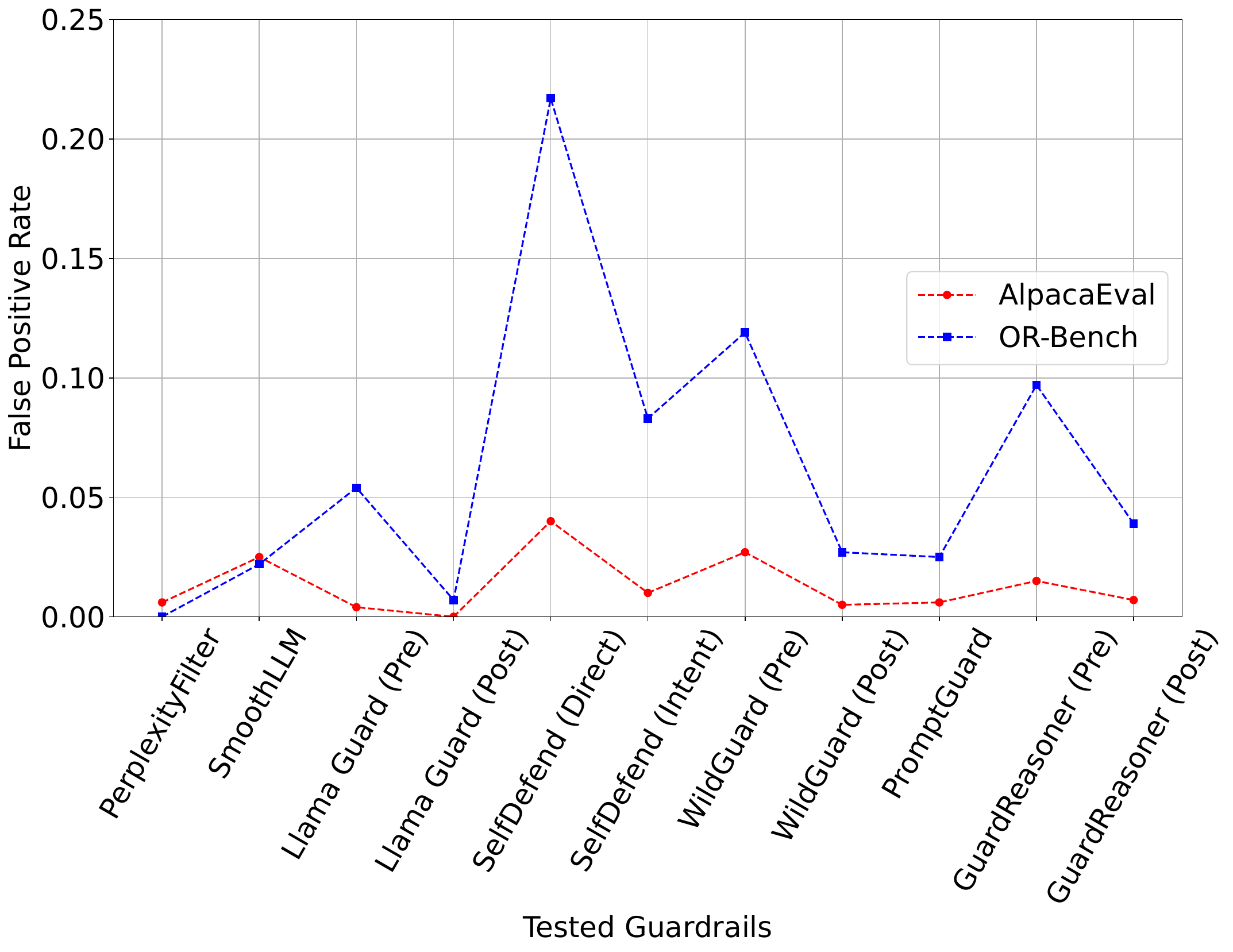}
	\caption{\revise{The utility (FPR) of guardrails on GPT-4-0125-Preview.}}
	\label{fig:fig_utility_gpt}
\end{figure}

\section{Evaluation Results on Vicuna-13b-v1.5}
\label{sec:vicuna_results}

We extend our comprehensive evaluation to another widely-used open-source model, Vicuna-13b-v1.5, to assess the generalization of different guardrails. The detailed results are presented in Table~\ref{tab:vicuna_results}.

First, a salient observation is that Vicuna-13b-v1.5 is considerably more susceptible to jailbreak attacks compared to Llama-3. This increased vulnerability is evident from the substantially higher Attack Success Rates (ASR) across almost all attack categories, indicating a weaker inherent safety alignment in Vicuna.

Second, we note a significant performance degradation for certain defenses when applied to Vicuna-13b-v1.5. For example, the efficacy of GradSafe and GradientCuff diminishes. GradientCuff, which showed marked effectiveness against the X-Teaming multi-turn attack on Llama-3, fails to maintain this advantage on Vicuna-13b-v1.5. This decline can be attributed to their nature as intra-processing guardrails, which heavily rely on the internal representations and alignment of the target LLM. Consequently, a less well-aligned model like Vicuna-13b-v1.5 compromises their defensive mechanism.

Despite these differences, we also observe consistent performance patterns. GuardReasoner (Pre) and GuardReasoner (Post) continue to exhibit state-of-the-art defense capabilities. GuardReasoner (Pre) achieves the best overall ASR of 0.156, while GuardReasoner (Post) records the best overall PGR of 0.192. This sustained excellence underscores that the robust defense mechanism of GuardReasoner is largely independent of the target LLM, positioning it as a more universally applicable and reliable guardrail.

\begin{table*}[t]
	\centering
    \caption{\revise{The ASR ($\downarrow$) / PGR ($\downarrow$) results for the target LLM (GPT-4-0125-Preview)
	.}
	}
	\vspace{-2.0ex}
	\resizebox{\textwidth}{!}{
		\revise{
		\begin{tabular}{l|c|cc|cc|cc|cc|c}
			\toprule
			\multirow{2}{*}{Guardrails} & Manual & \multicolumn{2}{c|}{Optimization-based} & \multicolumn{2}{c|}{Generation-based} &\multicolumn{2}{c|}{Implicit} &\multicolumn{2}{c|}{Multi-turn} & \multirow{2}{*}{Average}     \\ \cline{2-10}
			& IJP & GCG & AutoDAN & TAP & LLM-Fuzzer & DrAttack & MultiJail & ActorAttack & X-Teaming &   \\ \hline
            GPT-4-0125-Preview  &0.192/-&0.220/-&0.710/-&0.340/-&0.750/-&0.220/-&0.029/-&0.200/-&0.940/-&0.400/- \\
            PerplexityFilter     &0.192/1.000&0.000/0.060&0.710/1.000&0.260/1.000&0.720/1.000&0.220/1.000&0.029/1.000&0.200/1.000&0.930/1.000&0.362/0.896 \\
            SmoothLLM            &0.185/0.365&0.150/0.270&0.480/0.850&0.540/0.630&0.760/0.940&0.300/0.780&0.054/1.000&0.192/0.905&0.980/0.980&0.405/0.747 \\
            Llama Guard (Pre)    &0.100/0.563&0.130/0.460&0.540/0.770&0.250/0.510&0.660/0.840&0.220/0.850&0.029/0.952&0.200/0.998&0.940/0.990&0.341/0.770 \\
            Llama Guard (Post)   &0.110/0.110&0.180/0.180&0.530/0.530&0.250/0.250&0.720/0.720&0.220/0.220&0.029/0.029&0.200/0.200&0.950/0.950&0.354/0.354 \\
            SelfDefend (Direct)  &0.031/0.253&0.060/0.100&0.160/0.230&0.100/0.210&0.110/0.160&0.070/0.350&0.019/0.740&0.102/0.693&0.870/0.910&0.169/0.405 \\
            SelfDefend (Intent)  &0.039/0.310&0.040/0.090&0.080/0.160&0.100/0.190&0.100/0.180&0.060/0.140&0.025/0.568&0.128/0.768&0.910/0.930&0.165/0.371 \\
            WildGuard (Pre)      &0.004/0.033&0.000/0.020&0.020/0.020&0.030/0.060&0.010/0.010&0.140/0.530&0.029/0.797&0.127/0.768&0.810/0.880&\textbf{0.130}/0.346 \\
            WildGuard (Post)     &0.035/0.035&0.030/0.030&0.070/0.070&0.060/0.060&0.180/0.180&0.140/0.140&0.029/0.029&0.142/0.142&0.920/0.920&0.178/0.178 \\
            Prompt Guard         &0.000/0.000&0.020/0.050&0.410/0.630&0.310/0.960&0.000/0.000&0.220/0.970&0.029/1.000&0.195/0.993&0.910/1.000&0.233/0.623 \\
            GuardReasoner (Pre)  &0.000/0.009&0.000/0.000&0.030/0.030&0.030/0.040&0.050/0.070&0.080/0.290&0.016/0.349&0.113/0.740&0.890/0.960&0.134/0.276 \\
            GuardReasoner (Post) &0.021/0.021&0.030/0.030&0.040/0.040&0.080/0.080&0.240/0.240&0.100/0.100&0.029/0.029&0.110/0.110&0.930/0.930&0.176/\textbf{0.176} \\
			\bottomrule
		\end{tabular}
	}
	}
	\vspace{-2.0ex}
	\label{tab:gpt4_results}
\end{table*}


\revise{
\section{Evaluation Results on GPT-4-0125-Preview}
\label{sec:gpt4_results}
To further validate the generalizability of our findings, we evaluate guardrail performance on GPT-4-0125-Preview, a representative closed-source model. The comprehensive results are presented in Table~\ref{tab:gpt4_results}, Figure~\ref{fig:fig_delay_gpt}, \ref{fig:fig_memory_gpt}, and \ref{fig:fig_utility_gpt}.

Table~\ref{tab:gpt4_results} demonstrates that the defense performance metrics (ASR/PGR) on GPT-4 consistently align with those observed in Llama-based evaluations. Notably, PerplexityFilter achieves the highest average PGR, while GuardReasoner (Post) maintains the lowest. Furthermore, the multi-turn X-Teaming attack continues to exhibit exceptionally high ASR/PGR across all guardrails, reinforcing the findings from RQ3 regarding session-level defense limitations.

Figure~\ref{fig:fig_delay_gpt} reveals consistent efficiency patterns: GuardReasoner incurs substantial latency due to its reasoning overhead, and post-processing guardrails against jailbreaks universally introduce greater delay than their pre-processing counterparts, directly addressing RQ1. The pronounced latency of SmoothLLM stems from its defense mechanism requiring 10 perturbed prompt copies. This issue is further exacerbated by inefficient asynchronous API access to GPT-4 when compared to batched local inference.

Figure~\ref{fig:fig_memory_gpt} illustrates the GPU memory usage of different guardrails when applied to GPT-4. These GPU memory consumptions (in GB) for GPT-4 align with the conclusions drawn from Llama-3. Specifically: (1) GuardReasoner exhibits the highest memory footprint. (2) The memory overhead for SmoothLLM and PromptGuard is negligible. (3) LLM-based guardrails such as Llama Guard, SelfDefend, WildGuard, and GuardReasoner consume more GPU memory compared to rule-based guardrails (e.g., SmoothLLM). These have the same answer to RQ2.

The utility results (FPR) in Figure~\ref{fig:fig_utility_gpt} for GPT-4 corroborate the conclusions drawn from Llama-3. Specifically: (1) SelfDefend (Direct) exhibits the highest FPR on OR-Bench. (2) Token-level guardrails (SmoothLLM, SelfDefend (Direct)) demonstrate relatively pronounced FPRs. (3) Session-level guardrails (those with a ``Post" suffix) consistently achieve markedly lower FPRs than their sequence-level counterparts (those with a ``Pre" suffix). These findings consistently address RQ4.

These results demonstrate that our SEU framework and taxonomic insights generalize effectively to large-scale black-box models, affirming the robustness of our findings.

}

%% file: sec/review.tex
\newpage 


\section{Meta-Review}

The following meta-review was prepared by the program committee for the 2026
IEEE Symposium on Security and Privacy (S\&P) as part of the review process as
detailed in the call for papers.

\subsection{Summary}
This paper presents a defense-oriented systematization of knowledge on protecting large language models against jailbreak attacks. It proposes a taxonomy of jailbreak techniques and categorizes guardrail defenses along several key dimensions: the underlying technical paradigm, the granularity of protection, the degree of reactivity, their applicability across contexts, and the level of explainability they provide. In addition, the paper includes an evaluation of existing defenses, offering comparative insights into their effectiveness.

\subsection{Scientific Contributions}
\begin{itemize}
\item A systematization of defenses for LLMs
\end{itemize}

\subsection{Reasons for Acceptance}
\begin{enumerate}
\item This paper proposes a comprehensive taxonomy of LLM guardrails. The presentation is clear and has good coverage of existing work.
\item The paper also proposes a new evaluation framework that considers real-world trade-offs like latency, computational cost and usability.
\item The evaluation framework is demonstrated with various defenses and LLM models. This offers concrete benchmarks for comparing defenses.
\end{enumerate}

